\documentclass[sigconf]{acmart}

\AtBeginDocument{%
  \providecommand\BibTeX{{%
    \normalfont B\kern-0.5em{\scshape i\kern-0.25em b}\kern-0.8em\TeX}}}

\setcopyright{acmlicensed}
\copyrightyear{2024}
\acmYear{2024}
\setcopyright{rightsretained}
\acmConference[UIST '24]{The 37th Annual ACM Symposium on User Interface Software and Technology}{October 13--16, 2024}{Pittsburgh, PA, USA}
\acmBooktitle{The 37th Annual ACM Symposium on User Interface Software and Technology (UIST '24), October 13--16, 2024, Pittsburgh, PA, USA}
\acmDOI{10.1145/3654777.3676465}
\acmISBN{979-8-4007-0628-8/24/10}

\usepackage{caption}
\usepackage{subcaption}
\usepackage{cprotect}
\usepackage{lipsum}
\usepackage{svg}
\usepackage{fancyvrb}
\usepackage{tgcursor}
\usepackage{xfrac}
\usepackage{makecell}
\usepackage{color, colortbl}
\usepackage{tikz}
\usepackage[ruled]{algorithm2e}
\usepackage{listings}
\usepackage{wrapfig}
\usepackage{cleveref}
\usepackage{tikz}
\usepackage{framed}
\definecolor{block-gray}{gray}{0.95}
\newenvironment{codeblock}{%
  \MakeFramed{\advance\hsize-\width \FrameRestore}%
  \vspace{0.5ex}%
}{%
  \vspace{0.5ex}%
  \endMakeFramed%
}
\usepackage{multirow}
\usepackage{makecell}
\usepackage{hyperref}

\newcommand*\circled[1]{\tikz[baseline=(char.base)]{
    \node[shape=circle,draw,inner sep=1pt, fill=black, text=white] (char) {\textbf{\texttt{\footnotesize #1}}};}}

\newsavebox\spacewd
\savebox\spacewd{\texttt{ }}
\newenvironment{code}{\par\catcode32=\active \setlength{\parindent}{0pt}\small\ttfamily}{\par}
{
\catcode32=\active %
\gdef {\makebox[\wd\spacewd][l]{%
\phantom{\textcolor{white}{\fontfamily{lmtt}\selectfont\small\smash{\char32}}}}}%
}

\newcommand{\dash}{\,---\,}

\begin{document}

\title{Bluefish: Composing Diagrams with Declarative Relations}















\author{Josh Pollock}
\orcid{0000-0001-5141-0999}
\affiliation{%
  \institution{MIT CSAIL}
  \city{Cambridge}
  \state{MA}
  \country{USA}
}
\email{jopo@mit.edu}

\author{Catherine Mei}
\orcid{0009-0000-0334-0914}
\affiliation{%
  \institution{MIT CSAIL}
  \city{Cambridge}
  \state{MA}
  \country{USA}
}
\email{meic1212@mit.edu}

\author{Grace Huang}
\orcid{0009-0007-9840-065X}
\affiliation{%
  \institution{MIT CSAIL}
  \city{Cambridge}
  \state{MA}
  \country{USA}
}
\email{gracefh@mit.edu}

\author{Elliot Evans}
\orcid{0009-0005-9528-3828}
\affiliation{%
  \institution{Unaffiliated}
  \city{Ottawa}
  \state{Ontario}
  \country{Canada}
}
\email{vez@duck.com}

\author{Daniel Jackson}
\orcid{0000-0003-4864-078X}
\affiliation{%
  \institution{MIT CSAIL}
  \city{Cambridge}
  \state{MA}
  \country{USA}
}
\email{dnj@mit.edu}

\author{Arvind Satyanarayan}
\orcid{0000-0001-5564-635X}
\affiliation{%
  \institution{MIT CSAIL}
  \city{Cambridge}
  \state{MA}
  \country{USA}
}
\email{arvindsatya@mit.edu}

\renewcommand{\shortauthors}{Pollock et al.}

\begin{abstract}
Diagrams are essential tools for problem-solving and communication as they externalize conceptual structures using spatial relationships. But when picking a diagramming framework, users are faced with a dilemma. They can either use a highly expressive but low-level toolkit, whose API does not match their domain-specific concepts, or select a high-level typology, which offers a recognizable vocabulary but supports a limited range of diagrams. To address this gap, we introduce Bluefish: a diagramming framework inspired by component-based user interface (UI) libraries. Bluefish lets users create diagrams using \textit{relations}: declarative, composable, and extensible diagram fragments that relax the concept of a UI component. Unlike a component, a relation does not have sole ownership over its children nor does it need to fully specify their layout. To render diagrams, Bluefish extends a traditional tree-based scenegraph to a \textit{compound graph} that captures both hierarchical and adjacent relationships between nodes.
To evaluate our system, we construct a diverse example gallery covering many domains including mathematics, physics, computer science, and even cooking. We show that Bluefish's relations are effective declarative primitives for diagrams.
%
Bluefish is open source, and we aim to shape it into both a usable tool and a research platform.
\end{abstract}


\begin{teaserfigure}
  \includegraphics[width=\textwidth]{figures/teaser-v3.png}
  \caption{Diagrams built with the Bluefish language. These graphics run the gamut from computer science to physics to math, and are constructed with declarative, composable, extensible relations. From left to right: a quantum circuit equivalence~\cite{joy2020implementation}, topologies~\cite{Munkres1999-yi}, a Python Tutor diagram~\cite{pythontutor}, an Ohm parse tree~\cite{ohm}, and a physics pulley diagram~\cite{larkin1987diagram}.}
  \label{fig:teaser}
  \Description{Five diagrams created with the Bluefish language. From left to right: 1) a quantum circuit with two Hadamard gates and a controlled NOT gate, 2) a layered topology of concentric ellipses with labeled points, 3) a stack frame and heap tuple from a Python Tutor diagram, 4) a syntax parse tree of a mathematical expression, and 5) a physics diagram of a pulley system with labeled weights, ropes, and pulleys.}
\end{teaserfigure}


\maketitle

\section{Introduction}

Diagrams are essential to problem-solving and communication as they externalize conceptual structures as spatial relationships, thereby aiding recall, inference, and calculation~\cite{suwa2002external,norman2014things,larkin1987diagram}.
By representing information in new ways, the best diagrams unlock new ways of thinking about a problem domain\,---\,for example, by tracing events as a two-dimensional trajectory through space and time, Feynman diagrams opened ``new calculational vistas'' and quickly spread to many corners of modern physics~\cite{kaiser2005physics}.
Similarly, citing Dagonet, Latour argued that ``no scientific discipline exists without first inventing a visual and written language which allows it to break with its confusing past''~\cite{latour1986visualization,dagognet1973ecriture,franccois1969tableaux}.
Thus, scientific advances go hand-in-hand with novel diagrammatic notation. 

To produce diagrams, authors increasingly turn to programmatic frameworks as these tools enable data-driven diagramming, targeted rendering for different platforms (e.g., as vector graphics for web-based publishing or rasterized images for print-based media), and, more recently, automatic generation via large language models (LLMs). 
Existing frameworks, however, lie along a spectrum that trades off expressiveness for abstraction.
At one end are highly expressive toolkits, such as D3~\cite{d3} or p5.js~\cite{p5js}, that force authors to grapple with low-level concerns that are often orthogonal to the semantics of their domain-specific diagrams (e.g., manipulating the DOM or issuing Canvas drawing commands).
At the other end are high-level typologies, such as Mermaid~\cite{mermaid}, which offer a recognizable vocabulary of diagrams (flowcharts, sequence diagrams, etc.) but limit authors to only the available diagram types, with only a handful of customization options. 

To better balance between expressiveness and abstraction, we introduce \emph{Bluefish}, a diagramming framework inspired by modern component-based user interface (UI) toolkits such as React.
The basic building block of UI toolkits, the \emph{component}, offers authors several advantages: UIs can be specified \emph{declaratively}, in terms of what the interface should look like rather than how it should be laid out and rendered; components can be \emph{composed} together (e.g., by nesting them) to express custom UIs; and authors can \emph{extend} the specification language with custom components (e.g., to capture a recurring design pattern and make it reusable).
However, the component model imposes limitations when applied to authoring diagrams
(\Cref{sec:components}).
Components are assembled in tree-based hierarchies. But unlike UI elements, diagrammatic relationships frequently \textit{overlap}\dash{}a single element (e.g., a shape) may participate in many visual relationships simultaneously~(\Cref{fig:teaser}). 
These relationships cannot be easily expressed in a structure where an element can only have a single parent. As a result, in UI frameworks, diagram authors are forced to adopt low-level workarounds (e.g., manual bounding box calculations) that undo many of the advantages that components offer. 

In response, Bluefish relaxes the definition of a component to a \emph{relation} 
(\Cref{sec:lang-design}).
A relation, unlike a component, does not have sole ownership over its children nor does it need to fully specify their layout. 
Rather, a child element can be shared between multiple relations through \emph{scoped declarative references}, and its layout determined jointly by all parents.
With these changes, authors can smoothly trade locality for expressiveness 
(\Cref{sec:tradeoff})
\dash{}opting for a slightly more diffuse specification as it enables a more nimble prototyping process through the design space\dash{}without sacrificing the benefits of declarativity, composability, or extensibility. 

Authors construct Bluefish diagrams via the JSX syntax extension, which the language runtime compiles into a \emph{compound scenegraph} (\Cref{sec:relational-scenegraph})\,---\,an extension of a traditional tree-based scenegraph that captures both hierarchical and adjacency relationships between nodes. 
In contrast to traditional scenegraphs, compound scenegraphs introduce two challenges for layout: a node's layout may be specified by too few or too many parents. 
Thus,
to handle underconstrained systems, Bluefish \textit{lazily materializes} coordinate transforms~(\Cref{sec:materialize}) to ensure references can be properly resolved, even if the referent has not yet been fully positioned;
to handle overconstrained systems, Bluefish tracks \textit{bounding box ownership}~(\Cref{sec:ownership}) and notifies the user when relations conflict. 

To evaluate Bluefish, we developed a diverse gallery of example diagrams in collaboration with Elliot Evans, a professional creative coder\footnote{In recognition of his contribution, we include Evans as a co-author on this paper.}~(\Cref{sec:example-gallery}).
These diagrams span several domains including computer science, topology, physics, and cooking. We evaluate Bluefish's performance on three examples in this gallery and find that Bluefish's layout time scales linearly with the size of the scenegraph~(\Cref{sec:perf}). Additionally, we compare the relational design of Bluefish to the designs of Penrose~\cite{penrose} and Basalt~\cite{basalt}, two diagramming frameworks with different approaches to extending UI composition~(\Cref{sec:comparison}). Bluefish's component-inspired abstraction colocates data and display logic while Penrose, inspired by HTML and CSS, groups data and display logic separately. Bluefish uses declarative references that describe spatial relationships while Basalt uses low-level constraints. Finally, we reflect with Evans on Bluefish's abstraction design~(\Cref{sec:reflection}). We find that relaxing the component model provides a shallow learning curve for UI developers and that Bluefish's relational abstraction pushes specifications to be less hierarchical and more diffuse.

Our long-term goal is to make Bluefish both a usable tool and a research platform for investigating graphic representations from diagrams to documents to notation augmentations the way Vega-Lite has done for statistical graphics~\cite{vega-lite} and LLVM for compilers~\cite{lattner2004llvm}. To support this goal, we have released Bluefish as an open source project at \href{https://bluefishjs.org}{bluefishjs.org}, and we present several promising directions for future research and tool development~(\Cref{sec:discussion}).

\section{Related Work}

We first discuss the role of relations in diagrams, then we survey existing diagramming languages and environments, and finally we discuss approaches to layout.

\subsection{Relations and Diagrams}




Relations are central to diagramming.
According to James Clerk Maxwell,
a diagram is ``a figure drawn in such a manner that the geometrical relations between the parts of the figure illustrate relations between other objects''~\cite{maxwelldiagram}.
One salient kind of geometrical relation are \textit{Gestalt relations}~\cite{gestalt}, a collection of primitive visual relations that associate elements together. Some examples include distributing items with \textit{uniform density}; \textit{aligning} elements along a particular spatial axis;
containing elements in a \textit{common region}; and \textit{connecting} elements with lines or arrows.
Bluefish's relational standard library corresponds loosely to these relations:
%
\texttt{Distribute} to uniform density, \texttt{Align} to alignment, \texttt{Background} to common region, and \texttt{Arrow} and \texttt{Line} to connectedness. \texttt{Stack} uses a combination of alignment and uniform density.


Several researchers have formally analyzed diagrams in terms of their relations.  Richards extends Bertin's retinal variables (shape, color, size, etc.)~\cite{bertin1983semiology} with
Gestalt principles
to describe diagrams~\cite{Richards1984}. Building on this work, Engelhardt proposes a recursive language for diagram analysis~\cite{engelhardt}. The two have since collaborated on the VisDNA analysis grammar~\cite{engelhardt2021universal}. Larkin and Simon analyze diagrams by constructing formal relational data structures~\cite{larkin1987diagram}. These approaches have informed Bluefish's own relational formalism, but whereas these frameworks are designed for analysis, Bluefish allows a user to generate a diagram from a formal relational description.

\subsection{Diagramming Languages and Environments}


Direct manipulation editors like Figma, Omnigraffle, and tldraw allow users to align and distribute objects as well as attach arrows so they move when their attached objects are dragged. 
StickyLines makes alignment and distribution persistent, first-class objects that can be manipulated~\cite{ciolfi2016beyond}. Environments like Sketchpad~\cite{sutherland1963sketchpad}, TRIP~\cite{takahashi1998constraintTRIP}, GLIDE~\cite{ryall1997interactiveGLIDE}, Juno~\cite{nelson1985juno}, Dunnart~\cite{dwyer2009dunnart}, Delaunay~\cite{delaunayimpl}, Subform~\cite{subformappSubformDynamic}, and Charticulator~\cite{charticulator}, have integrated persistent relations in some form. Bluefish complements these approaches, because diagramming environments are typically based on a text representation that captures object state and sometimes constraints or relations between them. Text representations are typically better for defining and using abstractions than are direct manipulation environments. Text thus facilitates authoring data-driven diagrams and creating custom, reusable relation abstractions.


Some diagramming languages are restricted to just one or a few diagram types. GoTree supports tree diagrams~\cite{gotree}, and SetCoLa~\cite{setcola} and Graphviz~\cite{ellson2001graphviz} support node-link graphs using relations like alignment, spatial proximity, and connectedness. In contrast, Bluefish provides a relational standard library that is applicable across diagram types as well as mechanisms that allow users to customize and extend this set of relations.

General purpose diagramming languages provide various mechanisms for composition. 
Some UI frameworks, including Garnet~\cite{myers1995garnet}, Grow~\cite{barth1986objectGROW}, and Jetpack Compose~\cite{androidConstraintLayout}, extend the component model with an additional concept of a \textit{constraint}. Constraints are declarative equations or inequalities between variables that are solved by a constraint solver.  
Basalt~\cite{basalt} is a diagramming framework that does this as well.
Juno~\cite{nelson1985juno} and IDEAL~\cite{van1982highIDEAL}
allow users to specify reusable procedures that comprise collections of constraints. 
More recent diagramming systems have experimented with other abstractions. Haskell diagrams~\cite{yates2015diagrams} and Diagrammar~\cite{diagrammar} extend components with \textit{coordinate system modifiers}. For example, Haskell diagrams'
\texttt{align} function modifies the local origin of a component. Manim~\cite{manim} is a Python library for making animated diagrams. It extends components with \textit{imperative actions}. For example, Manim provides a \texttt{next\_to} method for shapes that, while similar to Bluefish's \texttt{Stack}, mutates objects.
This API is useful for making animations where objects change over time. Penrose~\cite{penrose}, a language for mathematical diagrams, 
uses constraints. But instead of organizing code with components, specifications are split across a \texttt{Substance} file and a \texttt{Style} file that are inspired by the split between HTML and CSS.
Bluefish draws inspiration from these systems' approaches to composition. However, rather than augmenting components with a new constraints concept, Bluefish relaxes the component model to relations. This provides a more consistent representation for relations than previous systems, which enables authors to more smoothly trade locality for expressiveness.

\subsection{Diagramming Layout Engines}

Laying out a diagram typically involves solving a system of constraints. Some engines use iterative methods that gradually converge on a solution such as Newton-Raphson~\cite{nelson1985juno}, force-direction~\cite{ryall1997interactiveGLIDE,setcola}, gradient descent~\cite{dwyer2009dunnart}, and L-BFGS~\cite{penrose}.
These methods can handle a diverse collection of constraints and can provide best-effort solutions when systems are unsatisfiable. Other systems pick a fixed constraint language and a specialized solver for it. TRIP~\cite{takahashi1998constraintTRIP}, IDEAL~\cite{van1982highIDEAL}, GoTree~\cite{gotree}, and Charticulator~\cite{charticulator}, for example, use linear programming. Basalt~\cite{basalt} lays out diagrams using SMT.

The above solvers are global: they solve an entire constraint system simultaneously. In contrast, local propagation methods flow information incrementally between constraints.
UI toolkits such as Garnet~\cite{myers1995garnet}, Grow~\cite{barth1986objectGROW}, and Apogee~\cite{henry1988usingAPOGEE} use this approach. 
In contrast to global solvers, local solvers are much simpler. Because they solve systems locally, it is easier to debug them when they go wrong. However, they can be less expressive. For example, a global solver can create an equilateral triangle from the constraints that three points are equidistant, but this cycle is not solvable with local propagation. More generally, global solvers tend to be better at continuous, geometric problems while local propagation solvers tend to be better at discrete problems. The UI layout engines in CSS~\cite{githubCsshoudinidraftscsslayoutapiEXPLAINERmdMain}, Android's Jetpack Compose~\cite{androidCustomLayouts}, and Apple's SwiftUI~\cite{appleComposeCustom} all employ local propagation strategies that are very similar to each other.
%
%
Just as we relax components to relations, we similarly generalize modern UI layout architectures.  

\section{Comparative Usage Scenarios}
\label{sec:components}

\begin{figure*}
    \centering
    \includegraphics[width=\textwidth]{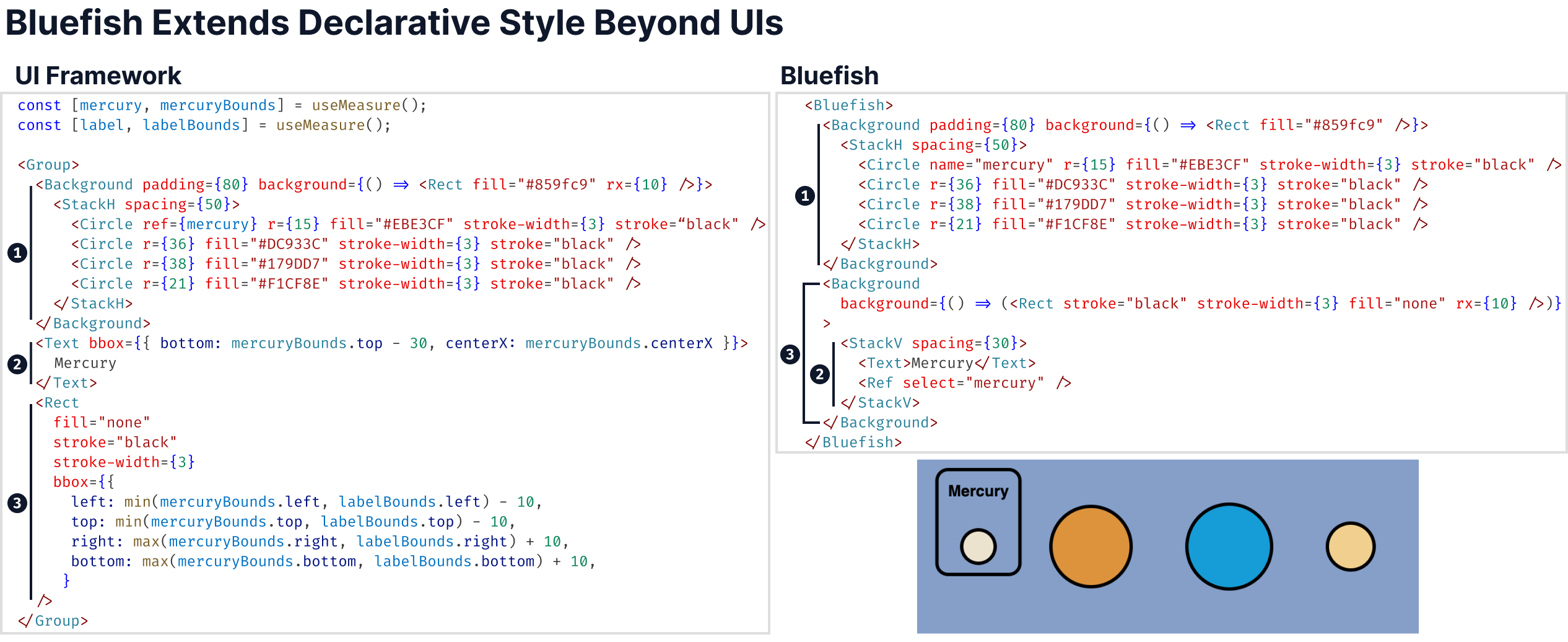}
    \caption{A comparison of specifying a simple diagram of the four terrestrial planets in a UI framework and in Bluefish. In both cases the user (1) makes a horizontal stack (\texttt{StackH}) of \texttt{Circle}s contained in a \texttt{Background}; (2) places a \texttt{Text} component above Mercury; and (3) surrounds the annotation and the planet with a background. In a UI framework, while (1) is declarative, (2) and (3) require error-prone, low-level bounding box computations. In contrast, all three steps are declarative in Bluefish.}
    \label{fig:ui-vs-bluefish}
    \Description{A side-by-side comparison of the code for an idealized UI framework and Bluefish to specify a simple planetary diagram. The top half shows the UI framework code and a horizontally arranged diagram of four colored circles representing planets against a rectangular background. The first circle is labeled Mercury and the label and circle are surrounded by a rectangle. The bottom half presents the Bluefish code, which is more succinct, and produces an identical diagram.}
\end{figure*}

To better motivate the need for Bluefish, and the design of its language, we begin with a walkthrough of how an author might construct a simple diagram (\Cref{fig:ui-vs-bluefish}) with common UI frameworks and compare what the process looks like with Bluefish instead.
This diagram depicts a row of the four terrestrial planets, with an annotation on Mercury.

\subsection{How UI Components Fail For Diagrams}


For this walkthrough, we imagine an \emph{idealized} UI framework. Since React relies on HTML and CSS to perform layout, we borrow component abstractions from SwiftUI and Jetpack Compose and express them with React's syntax for easier comparison with Bluefish. 

\circled{1} A user might start by creating a \texttt{StackH} (a horizontal stack, or row) of \texttt{Circle} marks and nest this inside a \texttt{Background} component:

\begin{codeblock}
\begin{code}
\definecolor{MineShaft}{RGB}{59,59,59}
\definecolor{Maroon}{RGB}{128,0,0}
\definecolor{JellyBean}{RGB}{38,127,153}
\definecolor{Red}{RGB}{229,0,0}
\definecolor{Black}{RGB}{0,0,0}
\definecolor{Blue}{RGB}{0,0,255}
\definecolor{Tamarillo}{RGB}{163,21,21}
\definecolor{Salem}{RGB}{9,134,88}
\textcolor{Maroon}{<}\textcolor{JellyBean}{Background }\textcolor{Red}{background}\textcolor{Black}{=}\textcolor{Blue}{\{}\textcolor{Black}{() }\textcolor{Blue}{=>}\textcolor{Black}{ ... }\textcolor{Blue}{\}}\textcolor{Maroon}{>}\\
\textcolor{MineShaft}{  }\textcolor{Maroon}{<}\textcolor{JellyBean}{StackH}\textcolor{MineShaft}{ }\textcolor{Red}{spacing}\textcolor{Black}{=}\textcolor{Blue}{\{}\textcolor{Salem}{50}\textcolor{Blue}{\}}\textcolor{Maroon}{>}\\
\textcolor{MineShaft}{    }\textcolor{Maroon}{<}\textcolor{JellyBean}{Circle}\textcolor{MineShaft}{ }\textcolor{Red}{r}\textcolor{Black}{=}\textcolor{Blue}{\{}\textcolor{Salem}{15}\textcolor{Blue}{\}}\textcolor{MineShaft}{ }\textcolor{Red}{fill}\textcolor{Black}{=}\textcolor{Blue}{\{}\textcolor{Tamarillo}{"\#EBE3CF"}\textcolor{Blue}{\}}\textcolor{MineShaft}{ }\textcolor{Black}{... }\textcolor{Maroon}{/>}\\
\textcolor{MineShaft}{    }\textcolor{Maroon}{<}\textcolor{JellyBean}{Circle}\textcolor{MineShaft}{ }\textcolor{Red}{r}\textcolor{Black}{=}\textcolor{Blue}{\{}\textcolor{Salem}{36}\textcolor{Blue}{\}}\textcolor{MineShaft}{ }\textcolor{Red}{fill}\textcolor{Black}{=}\textcolor{Blue}{\{}\textcolor{Tamarillo}{"\#DC933C"}\textcolor{Blue}{\}}\textcolor{MineShaft}{ }\textcolor{Black}{... }\textcolor{Maroon}{/>}\\
\textcolor{MineShaft}{    }\textcolor{Maroon}{<}\textcolor{JellyBean}{Circle}\textcolor{MineShaft}{ }\textcolor{Red}{r}\textcolor{Black}{=}\textcolor{Blue}{\{}\textcolor{Salem}{38}\textcolor{Blue}{\}}\textcolor{MineShaft}{ }\textcolor{Red}{fill}\textcolor{Black}{=}\textcolor{Blue}{\{}\textcolor{Tamarillo}{"\#179DD7"}\textcolor{Blue}{\}}\textcolor{MineShaft}{ }\textcolor{Black}{... }\textcolor{Maroon}{/>}\\
\textcolor{MineShaft}{    }\textcolor{Maroon}{<}\textcolor{JellyBean}{Circle}\textcolor{MineShaft}{ }\textcolor{Red}{r}\textcolor{Black}{=}\textcolor{Blue}{\{}\textcolor{Salem}{21}\textcolor{Blue}{\}}\textcolor{MineShaft}{ }\textcolor{Red}{fill}\textcolor{Black}{=}\textcolor{Blue}{\{}\textcolor{Tamarillo}{"\#F1CF8E"}\textcolor{Blue}{\}}\textcolor{MineShaft}{ }\textcolor{Black}{... }\textcolor{Maroon}{/>}\\
\textcolor{MineShaft}{  }\textcolor{Maroon}{</}\textcolor{JellyBean}{StackH}\textcolor{Maroon}{>}\\
\textcolor{Maroon}{</}\textcolor{JellyBean}{Background}\textcolor{Maroon}{>}
\end{code}
\end{codeblock}

\circled{2} But problems quickly arise when they try to annotate Mercury with some text. Ideally, the author should be able to place a \texttt{Text} component relative to the planet's position. That way if, for example, the \texttt{StackH}'s spacing or layout changed, the \texttt{Text} would move with it. However, the planet component is already contained within the \texttt{StackH} so it cannot participate directly in any other spatial relationships.\footnote{Note: One might be tempted to group the Mercury text and planet into a new component and use that in the \texttt{StackH}. But the Mercury text is longer than the planet, so grouping them together will affect the spacing between the planets.}

Situations like these can arise when specifying UIs as well (e.g., when placing a tooltip), so UI frameworks provide escape hatches to express more complicated layouts. A common escape hatch is a low-level layout or constraint API based on bounding boxes~\cite{androidConstraintLayout,androidCustomLayouts,appleComposeCustom}. For the purposes of demonstration, we present this as a hypothetical \texttt{useMeasure} hook, akin to those found in React. The \texttt{useMeasure} hook is a function that provides a reference that can be assigned to a component and a bounding box object that can be used to read and write dimensions of the referred component. Using this hook, the author could reference the bounding box of the Mercury circle and use it to position the text. They would first introduce a measure for Mercury, consisting of a reference and its bounding box:

\begin{codeblock}
\begin{code}
\definecolor{MineShaft}{RGB}{59,59,59}
\definecolor{Blue}{RGB}{0,0,255}
\definecolor{Lochmara}{RGB}{0,112,193}
\definecolor{Black}{RGB}{0,0,0}
\definecolor{Dallas}{RGB}{121,94,38}
\textcolor{Blue}{const}\textcolor{MineShaft}{ [}\textcolor{Lochmara}{mercury}\textcolor{MineShaft}{, }\textcolor{Lochmara}{mercuryBounds}\textcolor{MineShaft}{] }\textcolor{Black}{=}\textcolor{MineShaft}{ }\textcolor{Dallas}{useMeasure}\textcolor{MineShaft}{();}
\end{code}
\end{codeblock}

Then they would assign the \texttt{mercury} ref to the \texttt{Circle}:

\begin{codeblock}
\begin{code}
\definecolor{Maroon}{RGB}{128,0,0}
\definecolor{JellyBean}{RGB}{38,127,153}
\definecolor{Red}{RGB}{229,0,0}
\definecolor{Black}{RGB}{0,0,0}
\definecolor{Blue}{RGB}{0,0,255}
\definecolor{Lochmara}{RGB}{0,112,193}
\definecolor{Salem}{RGB}{9,134,88}
\textcolor{Maroon}{<}\textcolor{JellyBean}{Circle }\textcolor{Red}{ref}\textcolor{Black}{=}\textcolor{Blue}{\{}\textcolor{Lochmara}{mercury}\textcolor{Blue}{\} }\textcolor{Red}{r}\textcolor{Black}{=}\textcolor{Blue}{\{}\textcolor{Salem}{15}\textcolor{Blue}{\} }... \textcolor{Maroon}{/>}
\end{code}
\end{codeblock}

Finally, they would use this to compute the position of a new \texttt{Text} component:

\begin{codeblock}
\begin{code}
\definecolor{MineShaft}{RGB}{59,59,59}
\definecolor{Maroon}{RGB}{128,0,0}
\definecolor{JellyBean}{RGB}{38,127,153}
\definecolor{Red}{RGB}{229,0,0}
\definecolor{Black}{RGB}{0,0,0}
\definecolor{Blue}{RGB}{0,0,255}
\definecolor{NavyBlue}{RGB}{0,16,128}
\definecolor{Lochmara}{RGB}{0,112,193}
\definecolor{Salem}{RGB}{9,134,88}
\textcolor{Maroon}{<}\textcolor{JellyBean}{Text }\textcolor{Red}{bbox}\textcolor{Black}{=}\textcolor{Blue}{\{}\textcolor{Black}{\{ }\textcolor{NavyBlue}{bottom}\textcolor{NavyBlue}{:}\textcolor{Black}{ }\textcolor{Lochmara}{mercuryBounds}\textcolor{Black}{.}\textcolor{NavyBlue}{top}\textcolor{Black}{ - }\textcolor{Salem}{30}\textcolor{Black}{,}\\
            \textcolor{NavyBlue}{centerX}\textcolor{NavyBlue}{:}\textcolor{Black}{ }\textcolor{Lochmara}{mercuryBounds}\textcolor{Black}{.}\textcolor{NavyBlue}{centerX }\}\textcolor{Blue}{\}}\textcolor{Maroon}{>}\\
\textcolor{MineShaft}{  Mercury}\\
\textcolor{Maroon}{</}\textcolor{JellyBean}{Text}\textcolor{Maroon}{>}
\end{code}
\end{codeblock}

Unlike \circled{1}, this step requires the user to suddenly switch the level of abstraction they are working at: thinking explicitly about bounding boxes. Moreover, the user must remember that the y-axis of the coordinate system points down (they must use \texttt{-30} not \texttt{+30}) and that they have to offset the \texttt{bottom} of the \texttt{Text} component from the \texttt{top} of the \texttt{Circle}, and not vice versa.

\circled{3} These low-level escape hatches color the rest of the specification. Suppose the author now wants to place a \texttt{Background} behind the planet and the text to further emphasize their relationship. Again, since the \texttt{Circle} and the \texttt{Text} components have different parents, they cannot also be children of another \texttt{Background} component. Instead the author must again use bounding boxes. They first add a new measure:

\begin{codeblock}
\begin{code}
\definecolor{MineShaft}{RGB}{59,59,59}
\definecolor{Blue}{RGB}{0,0,255}
\definecolor{Lochmara}{RGB}{0,112,193}
\definecolor{Black}{RGB}{0,0,0}
\definecolor{Dallas}{RGB}{121,94,38}
\textcolor{Blue}{const}\textcolor{MineShaft}{ [}\textcolor{Lochmara}{label}\textcolor{MineShaft}{, }\textcolor{Lochmara}{labelBounds}\textcolor{MineShaft}{] }\textcolor{Black}{=}\textcolor{MineShaft}{ }\textcolor{Dallas}{useMeasure}\textcolor{MineShaft}{();}
\end{code}
\end{codeblock}

Then they assign it to the \texttt{Text}:

\begin{codeblock}
\begin{code}
\definecolor{MineShaft}{RGB}{59,59,59}
\definecolor{Maroon}{RGB}{128,0,0}
\definecolor{JellyBean}{RGB}{38,127,153}
\definecolor{Red}{RGB}{229,0,0}
\definecolor{Black}{RGB}{0,0,0}
\definecolor{Blue}{RGB}{0,0,255}
\definecolor{Lochmara}{RGB}{0,112,193}
\textcolor{Maroon}{<}\textcolor{JellyBean}{Text }\textcolor{Red}{ref}\textcolor{Black}{=}\textcolor{Blue}{\{}\textcolor{Lochmara}{label}\textcolor{Blue}{\}}\textcolor{Black}{ ...}\textcolor{Maroon}{>}\\
\textcolor{MineShaft}{  Mercury}\\
\textcolor{Maroon}{</}\textcolor{JellyBean}{Text}\textcolor{Maroon}{>}
\end{code}
\end{codeblock}

And finally they must perform another complicated bounding box computation to set the size of the background so that it contains both the planet and the label:

\begin{codeblock}
\begin{code}
\definecolor{MineShaft}{RGB}{59,59,59}
\definecolor{Maroon}{RGB}{128,0,0}
\definecolor{JellyBean}{RGB}{38,127,153}
\definecolor{Red}{RGB}{229,0,0}
\definecolor{Black}{RGB}{0,0,0}
\definecolor{Tamarillo}{RGB}{163,21,21}
\definecolor{Blue}{RGB}{0,0,255}
\definecolor{Salem}{RGB}{9,134,88}
\definecolor{NavyBlue}{RGB}{0,16,128}
\definecolor{Dallas}{RGB}{121,94,38}
\definecolor{Lochmara}{RGB}{0,112,193}
\textcolor{Maroon}{<}\textcolor{JellyBean}{Rect}\\
\textcolor{MineShaft}{  }\textcolor{Red}{fill}\textcolor{Black}{=}\textcolor{Tamarillo}{"none"}\\
\textcolor{MineShaft}{  }\textcolor{Red}{stroke}\textcolor{Black}{=}\textcolor{Tamarillo}{"black"}\\
\textcolor{MineShaft}{  }\textcolor{Red}{stroke-width}\textcolor{Black}{=}\textcolor{Blue}{\{}\textcolor{Salem}{3}\textcolor{Blue}{\}}\\
\textcolor{MineShaft}{  }\textcolor{Red}{bbox}\textcolor{Black}{=}\textcolor{Blue}{\{}\textcolor{Black}{\{}\\
\textcolor{Black}{    }\textcolor{NavyBlue}{left}\textcolor{NavyBlue}{:}\textcolor{Black}{ }\textcolor{Dallas}{min}\textcolor{Black}{(}\textcolor{Lochmara}{mercuryBounds}\textcolor{Black}{.}\textcolor{NavyBlue}{left}\textcolor{Black}{, }\textcolor{Black}{...),}\\
\textcolor{Black}{    }\textcolor{NavyBlue}{top}\textcolor{NavyBlue}{:}\textcolor{Black}{ }\textcolor{Dallas}{min}\textcolor{Black}{(}\textcolor{Lochmara}{mercuryBounds}\textcolor{Black}{.}\textcolor{NavyBlue}{top}\textcolor{Black}{, }\textcolor{Black}{...)}\textcolor{Black}{,}\\
\textcolor{Black}{    }\textcolor{NavyBlue}{right}\textcolor{NavyBlue}{:}\textcolor{Black}{ }\textcolor{Dallas}{max}\textcolor{Black}{(}\textcolor{Lochmara}{mercuryBounds}\textcolor{Black}{.}\textcolor{NavyBlue}{right}\textcolor{Black}{, }\textcolor{Black}{...)}\textcolor{Black}{,}\\
\textcolor{Black}{    }\textcolor{NavyBlue}{bottom}\textcolor{NavyBlue}{:}\textcolor{Black}{ }\textcolor{Dallas}{max}\textcolor{Black}{(}\textcolor{Lochmara}{mercuryBounds}\textcolor{Black}{.}\textcolor{NavyBlue}{bottom}\textcolor{Black}{, }\textcolor{Black}{...)}\textcolor{Black}{,}\\
\textcolor{Black}{  \}}\textcolor{Blue}{\}}\\
\textcolor{Maroon}{/>}
\end{code}
\end{codeblock}




To summarize, UI frameworks can prove to be quite brittle when expressing even very simple diagrams. This is because diagrams often contain relationships that break out of a tree-shaped component hierarchy. As a result users must resort to low-level escape hatches that do not closely match the semantics they want to express.

\subsection{Bluefish}



%
%
Now we consider how the same diagram is authored in Bluefish.

\circled{1} In Bluefish the user begins the same way as in the UI framework, except that their code is contained within a \texttt{Bluefish} tag:

\vspace{-1mm}
\begin{codeblock}
\begin{code}
\definecolor{MineShaft}{RGB}{59,59,59}
\definecolor{Maroon}{RGB}{128,0,0}
\definecolor{JellyBean}{RGB}{38,127,153}
\definecolor{Red}{RGB}{229,0,0}
\definecolor{Black}{RGB}{0,0,0}
\definecolor{Blue}{RGB}{0,0,255}
\definecolor{Salem}{RGB}{9,134,88}
\definecolor{Tamarillo}{RGB}{163,21,21}
\textcolor{Maroon}{<}\textcolor{JellyBean}{Bluefish}\textcolor{Maroon}{>}\\
\textcolor{MineShaft}{  }\textcolor{Maroon}{<}\textcolor{JellyBean}{Background}\textcolor{MineShaft}{ }\textcolor{Red}{background}\textcolor{Black}{=}\textcolor{Blue}{\{}\textcolor{Black}{() }\textcolor{Blue}{=>}\textcolor{Black}{ ...}\textcolor{Blue}{\}}\textcolor{Maroon}{>}\\
\textcolor{MineShaft}{    }\textcolor{Maroon}{<}\textcolor{JellyBean}{StackH}\textcolor{MineShaft}{ }\textcolor{Red}{spacing}\textcolor{Black}{=}\textcolor{Blue}{\{}\textcolor{Salem}{50}\textcolor{Blue}{\}}\textcolor{Maroon}{>}\\
\textcolor{MineShaft}{      }\textcolor{Maroon}{<}\textcolor{JellyBean}{Circle}\textcolor{MineShaft}{ }\textcolor{Red}{r}\textcolor{Black}{=}\textcolor{Blue}{\{}\textcolor{Salem}{15}\textcolor{Blue}{\}}\textcolor{MineShaft}{ }\textcolor{Red}{fill}\textcolor{Black}{=}\textcolor{Blue}{\{}\textcolor{Tamarillo}{"\#EBE3CF"}\textcolor{Blue}{\}}\textcolor{MineShaft}{ }\textcolor{Black}{... }\textcolor{Maroon}{/>}\\
\textcolor{MineShaft}{      }\textcolor{Maroon}{<}\textcolor{JellyBean}{Circle}\textcolor{MineShaft}{ }\textcolor{Red}{r}\textcolor{Black}{=}\textcolor{Blue}{\{}\textcolor{Salem}{36}\textcolor{Blue}{\}}\textcolor{MineShaft}{ }\textcolor{Red}{fill}\textcolor{Black}{=}\textcolor{Blue}{\{}\textcolor{Tamarillo}{"\#DC933C"}\textcolor{Blue}{\}}\textcolor{MineShaft}{ }\textcolor{Black}{... }\textcolor{Maroon}{/>}\\
\textcolor{MineShaft}{      }\textcolor{Maroon}{<}\textcolor{JellyBean}{Circle}\textcolor{MineShaft}{ }\textcolor{Red}{r}\textcolor{Black}{=}\textcolor{Blue}{\{}\textcolor{Salem}{38}\textcolor{Blue}{\}}\textcolor{MineShaft}{ }\textcolor{Red}{fill}\textcolor{Black}{=}\textcolor{Blue}{\{}\textcolor{Tamarillo}{"\#179DD7"}\textcolor{Blue}{\}}\textcolor{MineShaft}{ }\textcolor{Black}{... }\textcolor{Maroon}{/>}\\
\textcolor{MineShaft}{      }\textcolor{Maroon}{<}\textcolor{JellyBean}{Circle}\textcolor{MineShaft}{ }\textcolor{Red}{r}\textcolor{Black}{=}\textcolor{Blue}{\{}\textcolor{Salem}{21}\textcolor{Blue}{\}}\textcolor{MineShaft}{ }\textcolor{Red}{fill}\textcolor{Black}{=}\textcolor{Blue}{\{}\textcolor{Tamarillo}{"\#F1CF8E"}\textcolor{Blue}{\}}\textcolor{MineShaft}{ }\textcolor{Black}{... }\textcolor{Maroon}{/>}\\
\textcolor{MineShaft}{    }\textcolor{Maroon}{</}\textcolor{JellyBean}{StackH}\textcolor{Maroon}{>}\\
\textcolor{MineShaft}{  }\textcolor{Maroon}{</}\textcolor{JellyBean}{Background}\textcolor{Maroon}{>}\\
\textcolor{Maroon}{</}\textcolor{JellyBean}{Bluefish}\textcolor{Maroon}{>}
\end{code}
\end{codeblock}
\vspace{-1mm}

This tag demarcates the region of their specification that uses Bluefish's \emph{relations}\,---\,a relaxed definition of a UI component. While \texttt{Background} and \texttt{StackH} appear identically to their UI counterparts, in Bluefish we consider them to be relations and they can be used in more scenarios than before.

\circled{2} Rather than resort to bounding box computations to add the label, the user can use a relation instead. To do this, they first name the Mercury \texttt{Circle} so it can be referenced:

\vspace{-1mm}
\begin{codeblock}
\begin{code}
\definecolor{MineShaft}{RGB}{59,59,59}
\definecolor{Maroon}{RGB}{128,0,0}
\definecolor{JellyBean}{RGB}{38,127,153}
\definecolor{Red}{RGB}{229,0,0}
\definecolor{Black}{RGB}{0,0,0}
\definecolor{Tamarillo}{RGB}{163,21,21}
\definecolor{Blue}{RGB}{0,0,255}
\definecolor{Salem}{RGB}{9,134,88}
\textcolor{Maroon}{<}\textcolor{JellyBean}{Circle}\textcolor{MineShaft}{ }\textcolor{Red}{name}\textcolor{Black}{=}\textcolor{Tamarillo}{"mercury"}\textcolor{MineShaft}{ }\textcolor{Red}{r}\textcolor{Black}{=}\textcolor{Blue}{\{}\textcolor{Salem}{15}\textcolor{Blue}{\} }\textcolor{Black}{...}\textcolor{MineShaft}{ }\textcolor{Maroon}{/>}
\end{code}    
\end{codeblock}
\vspace{-1mm}

Then they write a \texttt{StackV} relation (\texttt{StackH}'s vertical counterpart) and select the existing planet element using a \texttt{Ref} component:

\vspace{-1mm}
\begin{codeblock}
\begin{code}
\definecolor{MineShaft}{RGB}{59,59,59}
\definecolor{JapaneseLaurel}{RGB}{128,128,128}
\definecolor{Maroon}{RGB}{128,0,0}
\definecolor{JellyBean}{RGB}{38,127,153}
\definecolor{Red}{RGB}{229,0,0}
\definecolor{Black}{RGB}{0,0,0}
\definecolor{Blue}{RGB}{0,0,255}
\definecolor{Salem}{RGB}{9,134,88}
\definecolor{Tamarillo}{RGB}{163,21,21}
\textcolor{MineShaft}{  }\textcolor{JapaneseLaurel}{</Background>}\\
\textcolor{MineShaft}{  }\textcolor{Maroon}{<}\textcolor{JellyBean}{StackV}\textcolor{MineShaft}{ }\textcolor{Red}{spacing}\textcolor{Black}{=}\textcolor{Blue}{\{}\textcolor{Salem}{30}\textcolor{Blue}{\}}\textcolor{Maroon}{>}\\
\textcolor{MineShaft}{    }\textcolor{Maroon}{<}\textcolor{JellyBean}{Text}\textcolor{Maroon}{>}\textcolor{MineShaft}{Mercury}\textcolor{Maroon}{</}\textcolor{JellyBean}{Text}\textcolor{Maroon}{>}\\
\textcolor{MineShaft}{    }\textcolor{Maroon}{<}\textcolor{JellyBean}{Ref}\textcolor{MineShaft}{ }\textcolor{Red}{select}\textcolor{Black}{=}\textcolor{Tamarillo}{"mercury"}\textcolor{MineShaft}{ }\textcolor{Maroon}{/>}\\
\textcolor{MineShaft}{  }\textcolor{Maroon}{</}\textcolor{JellyBean}{StackV}\textcolor{Maroon}{>}\\
\textcolor{JapaneseLaurel}{</Bluefish>}
\end{code}
\end{codeblock}
\vspace{-1mm}

Bluefish provides this special \texttt{Ref} component to allow relations to overlap\,---\,that is, for the same child element to participate in multiple relations simultaneously. Roughly speaking, \texttt{Ref} works as a proxy or stand-in for the element it selects. Since the \texttt{StackH} already placed the ``mercury'' \texttt{Circle}, the \texttt{StackV} treats it as a fixed element and positions the \texttt{Text} mark above it.

Compared to the explicit bounding box computations approach, using Bluefish's relations means the user does not have to remember low-level details like whether the label's bottom or top must be offset from the circle's top or bottom. Moreover, this specification is more declarative and has a closer mapping to the resultant diagram: as the \texttt{Text} mark is specified before the \texttt{Ref}, the label is vertically stacked \textit{above} the Mercury circle\dash{}a relationship they would have had to previously decode from low-level calculations. 

\circled{3} To place the rectangle behind the planet and the label, the user can wrap their new \texttt{StackV} in a \texttt{Background} relation:

\vspace{-1mm}
\begin{codeblock}
\begin{code}
\definecolor{MineShaft}{RGB}{59,59,59}
\definecolor{JapaneseLaurel}{RGB}{128,128,128}
\definecolor{Maroon}{RGB}{128,0,0}
\definecolor{JellyBean}{RGB}{38,127,153}
\definecolor{Red}{RGB}{229,0,0}
\definecolor{Black}{RGB}{0,0,0}
\definecolor{Blue}{RGB}{0,0,255}
\textcolor{MineShaft}{  }\textcolor{JapaneseLaurel}{</Background>}\\
\textcolor{MineShaft}{  }\textcolor{Maroon}{<}\textcolor{JellyBean}{Background}\textcolor{MineShaft}{ }\textcolor{Red}{background}\textcolor{Black}{=}\textcolor{Blue}{\{}\textcolor{Black}{() }\textcolor{Blue}{=>}\textcolor{Black}{ ...}\textcolor{Blue}{\}}\textcolor{Maroon}{>}\\
\textcolor{MineShaft}{    }\textcolor{JapaneseLaurel}{<StackV spacing=\{30\}>}\\
\textcolor{JapaneseLaurel}{      <Text>Mercury</Text>}\\
\textcolor{JapaneseLaurel}{      <Ref select="mercury" />}\\
\textcolor{JapaneseLaurel}{    </StackV>}\\
\textcolor{MineShaft}{  }\textcolor{Maroon}{</}\textcolor{JellyBean}{Background}\textcolor{Maroon}{>}\\
\textcolor{JapaneseLaurel}{</Bluefish>}
\end{code}
\end{codeblock}

Compared to the UI framework approach, Bluefish's relations allow the author to specify this diagram much more \textit{consistently}. A \texttt{Stack} is a \texttt{Stack} and a \texttt{Background} is a \texttt{Background} regardless of whether it is used in a conventional hierarchy or by referring to existing elements using \texttt{Ref}. This consistency extends the declarative nature of UI specifications to diagrams. As a result, compared to a UI framework, an author can create many more graphics using high-level APIs that closely match their intent.
\section{The Bluefish Language}
\label{sec:pg}

\begin{figure}
    \centering
    \includegraphics[width=0.5\textwidth]{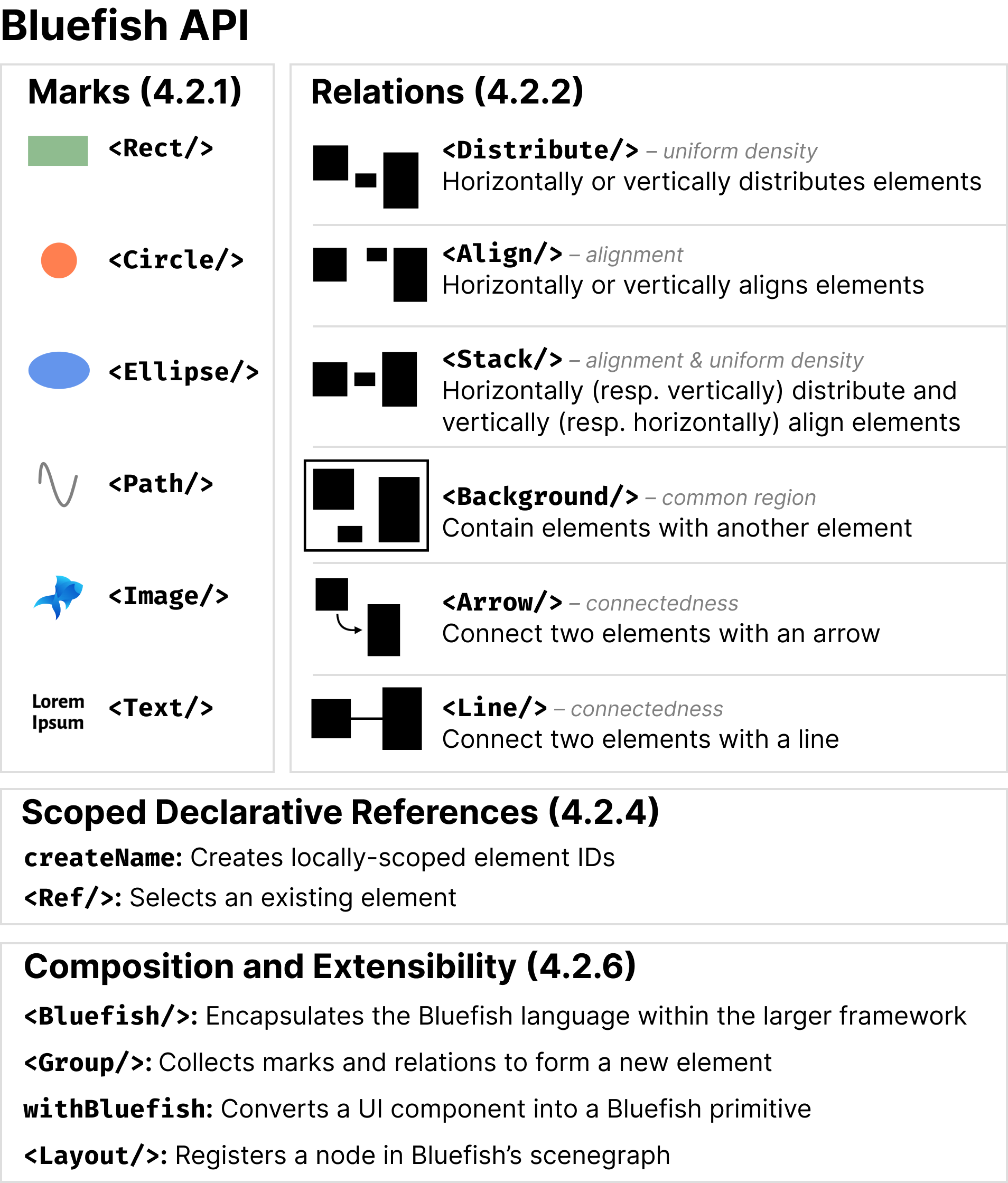}
    \caption{The Bluefish API comprises a standard library of marks and relations as well as a core set of language primitives. Bluefish relations are closely associated with Gestalt relations (listed in gray next to each tag). Scoped declarative references allow users to refer to existing elements. The \texttt{Group} relation, \texttt{withBluefish} function, and \texttt{Layout} component allow users to create new marks and relations.}
    \label{fig:api}
    \Description{A description of the Bluefish API, divided into four sections. Marks (4.2.1) are simple graphic elements: Rect, Circle, Ellipse, Path, Image, and Text. Relations (4.2.2) relate elements: Distribute horizontally or vertically distributes elements; Align horizontally or vertically aligns elements; Stack horizontally (resp. vertically) distributes and vertically (resp. horizontally) aligns elements; Background contains elements with another element; Arrow connects two elements with an arrow; Line connects two elements with a straight line. These relations are inspired by Gestalt relations: Distribute by uniform density, Align by alignment, Stack by alignment and uniform density, Background by common region, Arrow by connectedness, and Line by connectedness. Scoped Declarative References (4.2.4) are supported by createName, which creates locally-scoped element IDs, and the Ref tag, which selects an existing element. Composition and Extensibility (4.2.6) are supported by the Bluefish tag, which encapsulates the Bluefish language within a larger framework; the Group tag, which collects marks and relations to form a new element; the withBluefish function, which converts a UI component into a Bluefish primitive; and the Layout tag, which registers a node in Bluefish's scenegraph.}
\end{figure}

Bluefish is a domain-specific language (DSL) embedded in TypeScript comprising a standard library of basic marks and relations; scopes and references for overlapping relations; and helper functions and components for composing relations to \textit{create} new composite marks and relations. \Cref{fig:api} lists Bluefish's API.

The key innovation in Bluefish is its \textit{relation} abstraction. 
%
To address the limitations we describe in \Cref{sec:components}, relations relax the component model by allowing child elements to be shared across multiple parents via \emph{scoped declarative references}.
Moreover, a relation, unlike a component, can leave the sizes and positions of its children underspecified. 
As a result, the Bluefish relation concept allows users to smoothly trade locality for expressiveness~(\Cref{sec:tradeoff}).
By gradually making a specification more diffuse, a user can unlock spaces of atomic edits that they can then rapidly explore to prototype alternate diagram designs.

\subsection{Design Goals}
Motivated by the diagramming literature and by research on notational affordances, we identify three design goals to support expressive and flexible diagram authoring that UI components already exemplify. 
To these three, we add a fourth goal specific to diagrams that UI components poorly support.



\textbf{Declarative.} Declarative languages are popular across a range of domains (including web design via HTML/CSS, data querying with SQL, and data visualization using libraries such as Vega-Lite~\cite{vega-lite} or ggplot2~\cite{ggplot2}), because they decouple specification (the \emph{what}) from execution (the \emph{how}). As a result, authors are able to focus on their domain-specific concerns\dash{}in our case, expressing the semantics of their diagram\dash{}rather than contending with low-level computational and rendering considerations. Declarative specification is particularly important for diagramming as authors come from a variety of disciplinary backgrounds, with varying levels of expertise with reasoning about execution considerations.


\textbf{Composable.} In contrast to diagram typologies (e.g., Mermaid~\cite{mermaid}) which offer authors monolithic diagram types to pick between, we aim to achieve a greater expressive gamut by identifying a primitive set of building blocks that authors can combine together to achieve their desired output.
UI components support composition through \emph{nesting}: a component can be instantiated within another. 
This nesting structure is possible because components make few assumptions about the styling, layout, or state of their surrounding context. 
As a result, this compositional approach also allows authors to reason about their specification in a more localized manner\dash{}understanding one part at a time. Locality is especially important for authoring diagrams, which are often complex and non-hierarchical structures.


\textbf{Extensible.} While basic graphical shapes and elements\dash{}including rectangles, circles, lines, and text\dash{}provide the foundations of UIs and diagrams, they can present a greater \emph{articulatory distance}~\cite{hutchins1985direct} for expressing the semantics of a diagram than a more domain-specific set of primitives (e.g., ``pulleys,'' ``weights,'' and ``ropes'' for a physics diagram). As Ma'ayan \& Ni et al.~\cite{chi-diagrams} and the Cognitive Dimensions of Notations framework~\cite{blackwell2001cognitive} describe, it is important that authors have a specification language that has a correspondence~\cite{chi-diagrams} or close mapping~\cite{blackwell2001cognitive} to the vocabulary of their domain. 
However, it is impossible for language designers to anticipate every possible primitive for every potential domain. 
And, even if one could produce such a collection, it would impose an enormous maintenance burden on those designers. 
Thus, following UI components, Bluefish empowers authors to create domain-specific primitives.  
%

Finally, there is one additional design goal that a diagramming language must satisfy that UI components do not:

\textbf{Overlapping.} UI components can only relate to one-another via hierarchical nesting. This nesting partitions the visual plane into isolated sections that cannot easily communicate or be visually associated with each other except through a shared ancestor. Partitions help UI components achieve composability, because they can be reasoned about separately. But this trades off the expressiveness we need for diagramming. 
Diagram elements frequently crosscut a purely hierarchical structure\dash{}for example, the Mercury \texttt{Circle} in \Cref{fig:ui-vs-bluefish} participates in both a horizontal relationship with the other planetary circles and a vertical relationship with its text label.
Ideally a diagram author should be able to leverage locality when they can and expressiveness when they must.


\subsection{Language Design}
\label{sec:lang-design}


\subsubsection{Marks}
A mark is a basic visual element. Bluefish's mark standard library comprises \texttt{Rect}, \texttt{Circle}, \texttt{Ellipse}, \texttt{Path}, \texttt{Image}, and \texttt{Text}. Marks are thin wrappers around SVG primitives, except for \texttt{Text}. \texttt{Text} wraps visx's \texttt{Text} primitive, which provides better support for text layout than SVG's native \texttt{Text}.
A mark's position and size arguments are often omitted in a Bluefish specification, because they are determined by relations instead.

\subsubsection{Relations}
A relation is a visual arrangement of elements that conveys information about an abstract relationship between those elements (e.g., a line connecting two circles represents a chemical bond between two atoms).
In addition to marks, relations are the building blocks of diagrammatic representations~\cite{richards2002fundamental,engelhardt,larkin1987diagram,maxwelldiagram,tversky2001spatial}, and Bluefish's design reflects this.

Bluefish reifies the concept of a relation by relaxing UI components, the building blocks of user interfaces.
Relations are identical to UI components in many respects. For example, relations can contain zero or more children, be nested arbitrarily, and perform both rendering and layout. But relations relax the component model in two ways. First, whereas components' children are disjoint from other components' children, a relation may share children with other relations. This allows Bluefish elements to relate to other elements via multiple parent relations. Second, while a component must ensure its childrens' sizes and positions are fully determined, a relation can leave some unspecified for other relations to determine.
Together, these two relaxations allow Bluefish relations to overlap.

Bluefish's relations standard library is inspired by Gestalt relations~\cite{gestalt}. We provide relations that correspond to uniform density, alignment, common region, and connectedness~(\Cref{fig:api}). We selected Gestalt relations that are commonly found in UI toolkits and design tools. Toolkits like SwiftUI and Jetpack Compose provide components similar to \texttt{Stack} and \texttt{Background}, but they can only express \texttt{Distribute}, \texttt{Align}, \texttt{Arrow}, and \texttt{Line} relations indirectly through modifiers and bounding box calculations, because they do not support overlapping relations. In contrast, Bluefish supports all of these relations using the same abstraction. As a result, Bluefish's API more closely maps to Gestalt theory and allows users to more easily switch between different relations.

\subsubsection{Relations Are Expressed with JSX}
\label{sec:jsx}
We surface Bluefish's marks and relations through JSX, an extension to JavaScript popularized by the React library.
We map marks and relations to the two kinds of JSX tags: self-closing tags (e.g., \texttt{<Circle />}) for marks, and container tags (e.g., \texttt{<Arrow>...</Arrow>}) for relations. These tags instantiate elements with zero or non-zero children, respectively. JSX tags take attributes called \textit{props}. For example, \texttt{r} and \texttt{fill} are two of \texttt{<Circle />}'s props.

%
Our decision to represent relations using JSX instead of as props or vanilla functions has syntactic and semantic consequences.
Syntactically, modeling relations as components allows for a \emph{closeness of mapping}~\cite{blackwell2001cognitive}: a relation, which groups elements together, is defined by wrapping a container tag around participating elements, generalizing the notion of a grouping in other languages likely to be familiar to diagram authors including HTML (e.g., \texttt{<div>} tags) and SVG (i.e., \texttt{<g>} tags).
Semantically, representing relations as a single, consistent construct means they can be easily swapped for one another.
For example, many atomic edits like swapping a \texttt{Background} for an \texttt{Arrow} or replacing a \texttt{StackV} with an \texttt{Align} and a \texttt{Distribute} take advantage of this consistency~(\Cref{fig:viscosity}).

This representation stands in contrast to other diagramming frameworks. Haskell diagrams, for example, represents a stack using a function, but alignment using a coordinate transform. Similarly, Penrose represents a stack as collection of constraints and a background as a combination of a constraint and a mark. This makes atomic edits much more difficult.

\subsubsection{Scoped Declarative References (\texttt{<Ref />})}
\label{sec:ref}

To allow relations to share children, we provide a special \texttt{Ref} component that lets a user select an existing element to reuse as the child of another relation. None, some, or all of a relation's children may be \texttt{Ref}s.

A \texttt{Ref} works like a declarative query selector.
A user can reference an element by its \textit{name}. This name is either a globally defined string or scoped locally to the relation using the \texttt{createName} function. A user may also specify a \textit{path} of names.
To resolve a path selector, Bluefish traverses the path one-by-one, entering a relation each time and searching its local scope for the next named element.
Scopes encapsulate names so that changes to names in one relation definition cannot shadow names in another.

We considered using JavaScript's own variable bindings instead of a separate \texttt{Ref} component for specifying overlaps. 
However, we found that this interpretation of bindings competed with users' mental model of JSX: in JSX, using a component bound to a variable in multiple places creates different \textit{copies} of a component rather than \textit{referencing} it, as is needed with Bluefish. 
Moreover, using explicit \texttt{Ref}s simplifies the implementation of the system, because it allows us to construct a relational scenegraph within the confines of a tree-structured hierarchy. \Cref{sec:relational-scenegraph} explains this in more detail.
We leave to future work opportunities
to expand 
the expressiveness of how elements may be referenced (e.g., via XML query languages such as XPath~\cite{benedikt2009xpath,clark1999xml} and XQuery~\cite{boag2002xquery}, or by generalizing Cicero's \textit{specifiers}~\cite{cicero} and Atlas's \textit{find} function~\cite{atlas}).


\subsubsection{Relations Are Immutable}
Because of the hierarchical structure of a UI scenegraph, a component's layout behavior typically depends only on its props, its children, and its parent. As a result, a developer reading a UI codebase usually does not have to look at a component's siblings or cousins to determine the component's behavior. Since Bluefish allows siblings and cousins of a relation to also be its children, this introduces additional dependencies that could break the declarative nature of the component abstraction.
To reduce the impact of these non-local dependencies, Bluefish ensures that once some aspect of an elements's size or position, such as its width, has been set by a relation, no other relation may mutate it. As a result, whenever a diagram author sees a relation like \texttt{Align}, for example, a user can be confident that \texttt{Align}'s children are aligned regardless of other relations in the specification. We discuss how we enforce this property in \Cref{sec:layout}.

\subsubsection{Marks and Relations Are User-Extensible}
\label{sec:extensible-elements}

In addition to authoring diagrams with Bluefish's standard library, users can define new marks and relations in two ways. Firstly, since relations relax components, Bluefish inherits the compositional affordances of the UI framework model and JSX notation.
%
For example, we can write a custom \texttt{Planet} mark like so:
\begin{codeblock}
    \begin{code}
\definecolor{MineShaft}{RGB}{59,59,59}
\definecolor{Blue}{RGB}{0,0,255}
\definecolor{Lochmara}{RGB}{0,112,193}
\definecolor{Black}{RGB}{0,0,0}
\definecolor{Dallas}{RGB}{121,94,38}
\definecolor{NavyBlue}{RGB}{0,16,128}
\definecolor{Maroon}{RGB}{128,0,0}
\definecolor{JellyBean}{RGB}{38,127,153}
\definecolor{Red}{RGB}{229,0,0}
\textcolor{Blue}{const}\textcolor{MineShaft}{ }\textcolor{Lochmara}{Planet}\textcolor{MineShaft}{ }\textcolor{Black}{=}\textcolor{MineShaft}{ }\textcolor{Dallas}{withBluefish}\textcolor{MineShaft}{((}\textcolor{NavyBlue}{props}\textcolor{MineShaft}{) }\textcolor{Blue}{=>}\textcolor{MineShaft}{ (}\\
\textcolor{MineShaft}{  }\textcolor{Maroon}{<}\textcolor{JellyBean}{Circle}\textcolor{MineShaft}{ }\textcolor{Red}{r}\textcolor{Black}{=}\textcolor{Blue}{\{}\textcolor{NavyBlue}{props}\textcolor{Black}{.}\textcolor{NavyBlue}{radius}\textcolor{Blue}{\}}\textcolor{MineShaft}{ }\textcolor{Red}{fill}\textcolor{Black}{=}\textcolor{Blue}{\{}\textcolor{NavyBlue}{props}\textcolor{Black}{.}\textcolor{NavyBlue}{color}\textcolor{Blue}{\}}\textcolor{MineShaft}{}\textcolor{Maroon}{/>}\\
\textcolor{MineShaft}{));}
\end{code}
\end{codeblock}

The mark may then be used like a native tag: 
\begin{codeblock}
    \begin{code}
\definecolor{MineShaft}{RGB}{59,59,59}
\definecolor{Maroon}{RGB}{128,0,0}
\definecolor{Lochmara}{RGB}{0,112,193}
\definecolor{Black}{RGB}{0,0,0}
\definecolor{Blue}{RGB}{0,0,255}
\definecolor{Salem}{RGB}{9,134,88}
\definecolor{Tamarillo}{RGB}{163,21,21}
\textcolor{Maroon}{<}\textcolor{Lochmara}{Planet}\textcolor{MineShaft}{ }\textcolor{Lochmara}{radius}\textcolor{Black}{=}\textcolor{Blue}{\{}\textcolor{Salem}{15}\textcolor{Blue}{\}}\textcolor{MineShaft}{ }\textcolor{Lochmara}{color}\textcolor{Black}{=}\textcolor{Tamarillo}{"\#EBE3CF"}\textcolor{MineShaft}{ }\textcolor{Maroon}{/>}
\end{code}
\end{codeblock}

Any composition of marks and relations may be used as a custom mark, provided its elements have been completely sized and positioned relative to each other. For example, a user might rewrite \texttt{Planet} to place a \texttt{Background} around the planet as well:
\begin{codeblock}
\begin{code}
\definecolor{MineShaft}{RGB}{59,59,59}
\definecolor{Blue}{RGB}{0,0,255}
\definecolor{Lochmara}{RGB}{0,112,193}
\definecolor{Black}{RGB}{0,0,0}
\definecolor{Dallas}{RGB}{121,94,38}
\definecolor{NavyBlue}{RGB}{0,16,128}
\definecolor{Maroon}{RGB}{128,0,0}
\definecolor{JellyBean}{RGB}{38,127,153}
\definecolor{Red}{RGB}{229,0,0}
\textcolor{Blue}{const}\textcolor{MineShaft}{ }\textcolor{Lochmara}{Planet}\textcolor{MineShaft}{ }\textcolor{Black}{=}\textcolor{MineShaft}{ }\textcolor{Dallas}{withBluefish}\textcolor{MineShaft}{((}\textcolor{NavyBlue}{props}\textcolor{MineShaft}{) }\textcolor{Blue}{=>}\textcolor{MineShaft}{ (}\\
\textcolor{MineShaft}{}\textcolor{Maroon}{<}\textcolor{JellyBean}{Background}\textcolor{Maroon}{>}\\
\textcolor{MineShaft}{  }\textcolor{Maroon}{<}\textcolor{JellyBean}{Circle}\textcolor{MineShaft}{ }\textcolor{Red}{r}\textcolor{Black}{=}\textcolor{Blue}{\{}\textcolor{NavyBlue}{props}\textcolor{Black}{.}\textcolor{NavyBlue}{radius}\textcolor{Blue}{\}}\textcolor{MineShaft}{ }\textcolor{Red}{fill}\textcolor{Black}{=}\textcolor{Blue}{\{}\textcolor{NavyBlue}{props}\textcolor{Black}{.}\textcolor{NavyBlue}{color}\textcolor{Blue}{\}}\textcolor{MineShaft}{}\textcolor{Maroon}{/>}\\
\textcolor{MineShaft}{}\textcolor{Maroon}{</}\textcolor{JellyBean}{Background}\textcolor{Maroon}{>}\\
\textcolor{MineShaft}{));}
\end{code}
\end{codeblock}


When compositions of existing marks and relations is not enough, Bluefish allows users to author their own primitives with a low-level API. Inspired by the Jetpack Compose API~\cite{androidCustomLayouts}, primitive marks and relations are both described using a special \texttt{Layout} component that registers a node in Bluefish's scenegraph. In addition to taking a \texttt{name} (which may be provided implicitly by Bluefish), \texttt{Layout} requires a \texttt{layout} function that determines the bounding box and coordinate system of the element and has an opportunity to modify its children's bounding boxes and coordinate systems as well. \texttt{Layout} also requires a \texttt{paint} function that describes how the element should render given information about its bounding box and its children, which have already been rendered. Here is the basic structure for authoring a new primitive mark and a new primitive relation:

\begin{codeblock}
    \begin{code}
\definecolor{MineShaft}{RGB}{59,59,59}
\definecolor{Blue}{RGB}{0,0,255}
\definecolor{Lochmara}{RGB}{0,112,193}
\definecolor{Black}{RGB}{0,0,0}
\definecolor{Dallas}{RGB}{121,94,38}
\definecolor{NavyBlue}{RGB}{0,16,128}
\definecolor{Maroon}{RGB}{128,0,0}
\definecolor{ElectricViolet}{RGB}{175,0,219}
\definecolor{JellyBean}{RGB}{38,127,153}
\definecolor{Red}{RGB}{229,0,0}
\textcolor{Blue}{const}\textcolor{MineShaft}{ }\textcolor{Lochmara}{Rect}\textcolor{MineShaft}{ }\textcolor{Black}{=}\textcolor{MineShaft}{ }\textcolor{Dallas}{withBluefish}\textcolor{MineShaft}{((}\textcolor{NavyBlue}{props}\textcolor{MineShaft}{) }\textcolor{Blue}{=>}\textcolor{MineShaft}{ \{}\\
\textcolor{MineShaft}{ }\textcolor{Blue}{const}\textcolor{MineShaft}{ }\textcolor{Dallas}{layout}\textcolor{MineShaft}{ }\textcolor{Black}{=}\textcolor{MineShaft}{ () }\textcolor{Blue}{=>}\textcolor{MineShaft}{ \{ }\textcolor{Black}{...}\textcolor{MineShaft}{ \}}\\
\textcolor{MineShaft}{ }\textcolor{Blue}{const}\textcolor{MineShaft}{ }\textcolor{Dallas}{paint}\textcolor{MineShaft}{ }\textcolor{Black}{=}\textcolor{MineShaft}{ (}\textcolor{NavyBlue}{paintProps}\textcolor{MineShaft}{) }\textcolor{Blue}{=>}\textcolor{MineShaft}{ }\textcolor{Maroon}{<}\textcolor{Maroon}{rect}\textcolor{MineShaft}{ }\textcolor{Black}{...}\textcolor{MineShaft}{ }\textcolor{Maroon}{/>}\\
\textcolor{MineShaft}{ }\textcolor{ElectricViolet}{return}\textcolor{MineShaft}{ }\textcolor{Maroon}{<}\textcolor{JellyBean}{Layout}\textcolor{MineShaft}{ }\textcolor{Red}{layout}\textcolor{Black}{=}\textcolor{Blue}{\{}\textcolor{NavyBlue}{layout}\textcolor{Blue}{\}}\textcolor{MineShaft}{ }\textcolor{Red}{paint}\textcolor{Black}{=}\textcolor{Blue}{\{}\textcolor{Dallas}{paint}\textcolor{Blue}{\}}\textcolor{MineShaft}{}\textcolor{Maroon}{/>}\\
\textcolor{MineShaft}{\})}
\end{code}
\end{codeblock}

\begin{codeblock}
\begin{code}
\definecolor{MineShaft}{RGB}{59,59,59}
\definecolor{Blue}{RGB}{0,0,255}
\definecolor{Lochmara}{RGB}{0,112,193}
\definecolor{Black}{RGB}{0,0,0}
\definecolor{Dallas}{RGB}{121,94,38}
\definecolor{NavyBlue}{RGB}{0,16,128}
\definecolor{Maroon}{RGB}{128,0,0}
\definecolor{ElectricViolet}{RGB}{175,0,219}
\definecolor{JellyBean}{RGB}{38,127,153}
\definecolor{Red}{RGB}{229,0,0}
\textcolor{Blue}{const}\textcolor{MineShaft}{ }\textcolor{Lochmara}{Align}\textcolor{MineShaft}{ }\textcolor{Black}{=}\textcolor{MineShaft}{ }\textcolor{Dallas}{withBluefish}\textcolor{MineShaft}{((}\textcolor{NavyBlue}{props}\textcolor{MineShaft}{) }\textcolor{Blue}{=>}\textcolor{MineShaft}{ \{}\\
\textcolor{MineShaft}{  }\textcolor{Blue}{const}\textcolor{MineShaft}{ }\textcolor{Dallas}{layout}\textcolor{MineShaft}{ }\textcolor{Black}{=}\textcolor{MineShaft}{ (}\textcolor{NavyBlue}{childNodes}\textcolor{MineShaft}{) }\textcolor{Blue}{=>}\textcolor{MineShaft}{ \{ }\textcolor{Black}{...}\textcolor{MineShaft}{ \}}\\
\textcolor{MineShaft}{  }\textcolor{Blue}{const}\textcolor{MineShaft}{ }\textcolor{Dallas}{paint}\textcolor{MineShaft}{ }\textcolor{Black}{=}\textcolor{MineShaft}{ (}\textcolor{NavyBlue}{paintProps}\textcolor{MineShaft}{) }\textcolor{Blue}{=>}\textcolor{MineShaft}{ (}\textcolor{Maroon}{<}\textcolor{Maroon}{g}\textcolor{MineShaft}{ }\textcolor{Black}{...}\textcolor{MineShaft}{ }\textcolor{Maroon}{>}\\
\textcolor{MineShaft}{    }\textcolor{Blue}{\{}\textcolor{NavyBlue}{paintProps}\textcolor{Black}{.}\textcolor{NavyBlue}{children}\textcolor{Blue}{\}}\\
\textcolor{MineShaft}{  }\textcolor{Maroon}{</}\textcolor{Maroon}{g}\textcolor{Maroon}{>}\textcolor{MineShaft}{)}\\
\textcolor{MineShaft}{  }\textcolor{ElectricViolet}{return}\textcolor{MineShaft}{ (}\\
\textcolor{MineShaft}{    }\textcolor{Maroon}{<}\textcolor{JellyBean}{Layout}\textcolor{MineShaft}{ }\textcolor{Red}{layout}\textcolor{Black}{=}\textcolor{Blue}{\{}\textcolor{NavyBlue}{layout}\textcolor{Blue}{\}}\textcolor{MineShaft}{ }\textcolor{Red}{paint}\textcolor{Black}{=}\textcolor{Blue}{\{}\textcolor{Dallas}{paint}\textcolor{Blue}{\}}\textcolor{Maroon}{>}\\
\textcolor{MineShaft}{      }\textcolor{Blue}{\{}\textcolor{NavyBlue}{props}\textcolor{Black}{.}\textcolor{NavyBlue}{children}\textcolor{Blue}{\}}\\
\textcolor{MineShaft}{    }\textcolor{Maroon}{</}\textcolor{JellyBean}{Layout}\textcolor{Maroon}{>}\\
\textcolor{MineShaft}{  );}\\
\textcolor{MineShaft}{\})}
\end{code}
\end{codeblock}

Bluefish's standard library is written using this API. As a result, it is fully customizable and extensible from user space. We discuss how \texttt{layout} functions work in \Cref{sec:layout}.


\subsection{Design Implication: Smoothly Trading Locality for Expressiveness}
\label{sec:tradeoff}

\begin{figure*}
    \centering
    \includegraphics[width=0.8\textwidth]{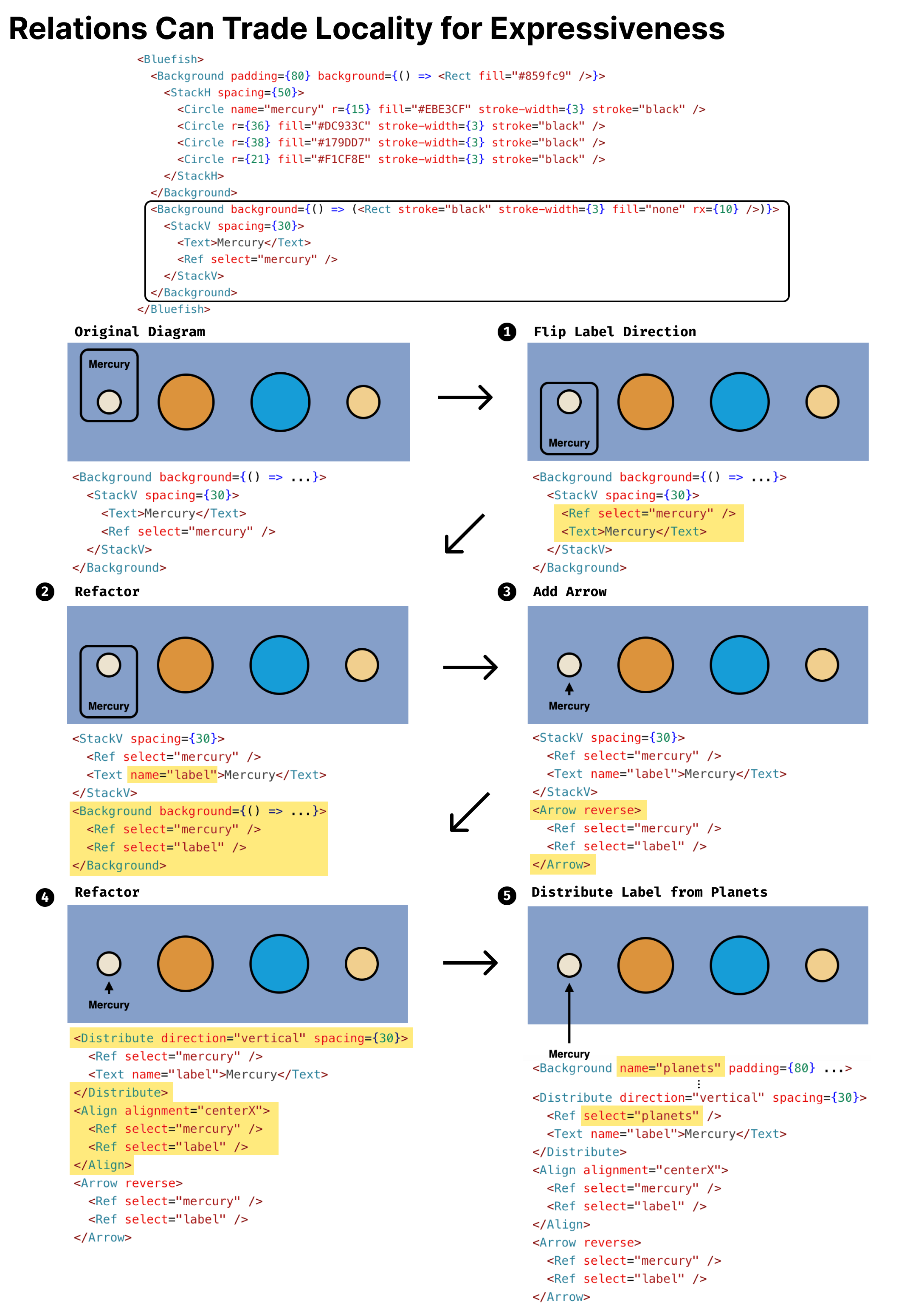}
    \caption{Relations can trade locality for expressiveness. (1) With the original spec, one can flip the direction of the label. (2-3) After denesting the \texttt{Background} and \texttt{StackV} relations one can replace the \texttt{Background} with an \texttt{Arrow}. (4-5) After breaking up \texttt{StackV} into \texttt{Align} and \texttt{Distribute}, one can space the label relative to the planets background while keeping it aligned with Mercury.}
    \label{fig:viscosity}
    \Description{A graphic describing how Bluefish can trade locality for expressiveness via progressive modifications and refactorings. (1) The label is shifted from above the planet to below it. (2) The code is denested without changing the diagram. (3) The label-planet relation is switched from a Background to an Arrow. (4) The StackV is split apart without changing the diagram. (5) The label is moved outside the Background by retargeting the Distribute relation.}
\end{figure*}

In addition to extending the declarative component model to more complex graphics, Bluefish allows a user more flexibility to trade locality for expressiveness. Specifically one can make a specification more diffuse by \textit{denesting} or \textit{breaking up} relations. These processes preserve the output diagram while affording new atomic ways to modify the specification. Consider \Cref{fig:viscosity}.

\circled{1} Starting with the specification from~\Cref{fig:ui-vs-bluefish}, one can already make some atomic edits to explore alternative designs. For example, by swapping the order of the \texttt{StackV}'s children, one can move the label below the planet.
%
%
\circled{2} By \textit{denesting} the \texttt{Background} and \texttt{StackV} relations, one can make the specification a little more diffuse. This can be accomplished by naming the \texttt{Text} mark, moving the \texttt{Background} so it is adjacent to the \texttt{StackV} instead of containing it, and making \texttt{Background}'s children \texttt{Ref}s to \texttt{StackV}'s children.
\circled{3} This results in a more verbose specification than in \circled{1}. But the advantage is that now the \texttt{Background} can be replaced with another relation like \texttt{Arrow} simply by swapping the tag.
%
%
\circled{4} Finally, for even more expressiveness, one can split \texttt{StackV} into two more primitive relations. \texttt{StackV} is a compound relation that horizontally \textit{aligns} its children and vertically \textit{distributes} them. Bluefish provides \texttt{Align} and \texttt{Distribute} so that a user can specify these relations individually.
\circled{5} Splitting \texttt{StackV} allows one to retarget \texttt{Distribute} at different children while keeping \texttt{Align} fixed. For example one can place the label outside the planets \texttt{Background} as follows. First, label the \texttt{Background}, ``planets'', and then change the first child of the \texttt{Distribute} to select it. This positions the label so that it is still horizontally aligned with Mercury but vertically spaced relative to the planets.

Notice \texttt{Align} and \texttt{Distribute} cannot be expressed as components in UI frameworks. This is because they only control their childrens' positions on one axis and so those children must have more than one parent to be fully positioned. In SwiftUI and Jetpack Compose, alignment is available as a \textit{guide} argument to components like \texttt{HStack} or as a \textit{modifier} on individual elements.
Compose also exposes align and distribute \textit{constraints} in special \texttt{ConstraintLayout} components.
To summarize, Bluefish's relation model allows one to smoothly trade locality for expressiveness. One can make a specification more diffuse by denesting relations and breaking them apart. By doing so, one gains more opportunities to make atomic edits.

\section{The Bluefish Relational Scenegraph}
\label{sec:relational-scenegraph}


Bluefish is a implemented in SolidJS, a reactive UI framework. Solid provides a JSX component abstraction, signal library, and renderer for Bluefish.
We maintain a separate scenegraph and provide a custom layout engine for this scenegraph.
When a user composes Bluefish marks and relations, the language runtime compiles this specification to a \emph{relational scenegraph}: a data structure used to resolve references between elements and compute layout. 
Critically, in contrast to tree-based scenegraphs that are standard in UI and visualization toolkits, Bluefish's relational scenegraph is an instance of a \emph{compound graph}: a data structure that maintains the hierarchical information of traditional scenegraphs while also encoding adjacency relationships between nodes~(\Cref{sec:ref}).
To compute layout, we adopt the principle of \emph{conservative extension}~\cite{felleisen1991expressive} such that when a relational scenegraph is purely hierarchical its layout behavior is indistinguishable from the behavior of a tree-based scenegraph. This principle allows us to extend the benefits of UI layout runtimes to Bluefish.

\begin{figure*}[t]
    \centering
\includegraphics[width=0.9\textwidth]{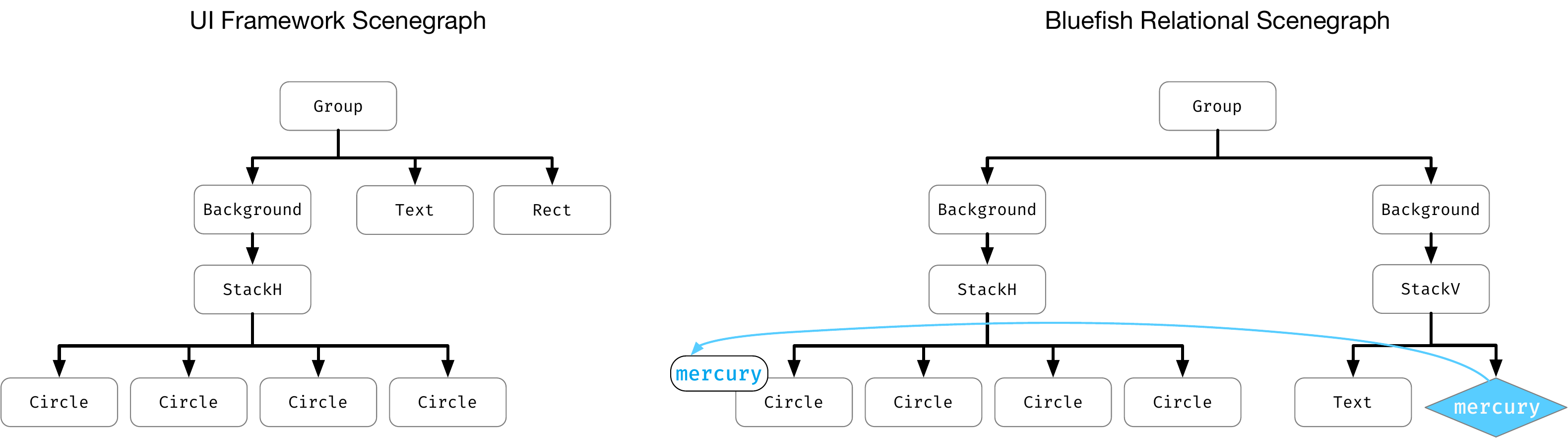}
    \caption{The tree-structured and relational scenegraphs corresponding to \autoref{fig:ui-vs-bluefish}.
    Notice that Bluefish's scenegraph retains more information than a tree-structured scenegraph. Bluefish's scenegraph represents the \texttt{StackV} and \texttt{Background} relations between the label and the planet.
    In the traditional scenegraph the \texttt{StackV} relation between the label and the planet is missing, and the \texttt{Background} relation has been reduced to a \texttt{Rect} component.}
    \label{fig:relational_scenegraph}
    \Description{A comparison of the UI scenegraph and Bluefish scenegraph structures for the code in Figure 2. The UI scenegraph is as follows:
Group
- Background
-- StackH
--- Circle
--- Circle
--- Circle
--- Circle
- Text
- Rect
The Bluefish scenegraph is as follows:
Group
- Background
-- StackH
--- Circle [mercury]
--- Circle
--- Circle
--- Circle
- Background
-- StackV
--- Text
--- Ref mercury

Figure 6: A diagram of how the lazy materialization algorithm initializes transforms in the scenegraph.
1. Before StackV lays out.
Group (x ?, y ?)
- Rect (x ?, y ?) [a]
- Rect (x ?, y ?) [b]
- StackV (x ?, y ?)
-- Ref a
-- Ref b
2. StackV attempts to set a's x position.
3. We lazily materialize StackV's x transform to ensure StackV is horizontally fixed relative to a.
Group (x ?, y ?)
- Rect (x -5, y ?)
- Rect (x ?, y ?)
- StackV (x 0, y ?)
-- Ref a
-- Ref b
4. StackV sets b's x position. This does not require additional materialization.
Group (x ?, y ?)
- Rect (x -5, y ?)
- Rect (x -15, y ?)
- StackV (x 0, y ?)
-- Ref a
-- Ref b
5. We materialize StackV's y transform when StackV attempts to set a's y position
Group (x ?, y ?)
- Rect (x -5, y 0)
- Rect (x -15, y 50)
- StackV (x 0, y 0)
-- Ref a
-- Ref b}
\end{figure*}

\subsection{Adapting a Compound Graph Structure}
\label{sec:compound-graph}

Compound graphs have been explored in research on graph drawing~\cite{sugiyama2002graph} and hierarchical edge bundling~\cite{holten2006hierarchical}. 
They encode not only \emph{hierarchical} relationships between nodes (i.e., parent-child) but also allow for non-hierarchical relationships called \emph{adjacencies}.
In Bluefish, we instantiate a compound graph as follows:

\emph{Nodes.} 
Each node in the scenegraph corresponds to a \texttt{Layout} or \texttt{Ref} tag instantiated in JSX. 
\texttt{Layout} nodes hold information necessary for rendering the corresponding element (e.g., visual styles) as well as computing layout. 
Layout information includes a partially defined bounding box
and any transformations needed to position and size this node based on higher-level nodes. 

\emph{Hierarchy and Adjacencies.}
Nodes are assembled into a hierarchy following the nesting structured established by the JSX specification. 
An adjacency relation is established for every \texttt{Ref} element: a node is instantiated in the hierarchy for the \texttt{Ref}, and it links to the referenced node as an adjacency.
As a result, and unlike general compound graphs where adjacencies can connect any pair of nodes, adjacencies in Bluefish always originate at leaf nodes (i.e., \texttt{Ref}s are self-closing tags rather than block or container tags).
\autoref{fig:relational_scenegraph} depicts the scenegraph for the UI specification and the relational scenegraph for the Bluefish specification from~\autoref{fig:ui-vs-bluefish}.

\subsection{Computing Layout by Conservatively Extending UI Tree-Based Local Propagation}
\label{sec:layout}


Bluefish extends the layout architecture adopted by modern UI layout engines including those underlying
CSS~\cite{githubCsshoudinidraftscsslayoutapiEXPLAINERmdMain}, SwiftUI~\cite{appleBuildingCustom,appleComposeCustom}, and Jetpack Compose~\cite{androidCustomLayouts}. This architecture is a form of tree-based \textit{local propagation}~\cite{steele1980definition}. 
Local propagation has a storied history in UI toolkits~\cite{sannella1993multi,sannella1994skyblue,borning1996indigo,myers1991graphical,vander2001lessons,myers1995garnet,barth1986objectGROW,henry1988usingAPOGEE} and is straightforward to implement in reactive dataflow runtimes.


UI layout needs to be fast yet support different, specialized algorithms like flex layout, line-breaking, and grids. To balance performance and expressiveness, UI layouts execute in one pass over the scenegraph, and each node can contain arbitrary code.
Each node in the scenegraph has an associated layout algorithm, and layout commences at the scenegraph root. 
When a node's layout algorithm is evaluated, it invokes the algorithms of its children by proposing a width and height for each child. 
Once the children are laid out, they return their actual sizes and the parent may place each child in its local coordinate system.
This local information-passing approach can express many kinds of layouts. For example, to implement flex layout each child may optionally specify its flex factor. During layout, the parent flex node can read its children's flex factors and distribute its free space proportionally to each child.

Instead of local propagation, many visualization and diagramming frameworks use a different strategy, a \textit{global solver}~\cite{nelson1985juno,ryall1997interactiveGLIDE,setcola,penrose,dwyer2009dunnart,takahashi1998constraintTRIP,van1982highIDEAL,gotree,charticulator,basalt} such as gradient descent, linear programming, or SMT. Whereas a local solver specifies how values flow through a constraint graph, a global solver specifies a constraint language that all layout constraints must be written in such as differential programs for gradient descent, linear inequalities and an objective function for linear programming, or quantifier-free non-linear real arithmetic for SMT.
Because global solvers solve all constraints simultaneously, they can tackle very complex layout problems that cannot be solved by local propagation. 
Indeed, we implemented an early version of Bluefish using the Cassowary linear programming solver~\cite{badros2001cassowary}. 
However, in doing so, we identified a series of tradeoffs at odds with our design goals of composability and extensibility.
First, a global solver increases viscosity for diagram authors: it can be difficult to localize layout bugs because the solver reasons about all constraints at once and a node's layout can, by design, be a function of a highly non-local set of inputs. 
Second, while global solvers increase expressiveness by supporting a larger class of layout problems, they actually limit extensibility: common domain-specific algorithms for domains like trees~\cite{nonlayered-tidy} and graphs~\cite{sugiyama} rely on custom imperative code that cannot be easily translated to or integrated with a global solver's constraint language. 
In contrast, Bluefish layout problems can be debugged more easily as layout information only flows locally. 
Moreover, Bluefish is able to integrate any external layout algorithm simply by instantiating it as a node in the scenegraph. These benefits are extended directly from UI layout architectures.

Local propagation does present some limitations\dash{}namely, it does not provide special support for continuous optimization problems or complicated simultaneous constraints. Many International Math Olympiad geometry problems, for example, can only be drawn by solving a circular system of geometry constraints~\cite{krueger2021automatically}. Diagrams involving knots are well-suited to gradient-descent schemes~\cite{yu2021repulsive}. Nevertheless, such domain-specific solvers could be embedded as special nodes in the Bluefish architecture. In this way, Bluefish serves more as a layout fabric than a solver itself. It is concerned with the interface between nodes more than the language those nodes' layouts are written in.

\subsubsection{A Running Example: Equivalent \texttt{StackV} Specifications}

\begin{algorithm}
\caption{The layout algorithm for \texttt{StackV}}
\label{alg:col}
\DontPrintSemicolon
\SetArgSty{upshape}
\KwData{alignment, spacing}
\nlset{XY}\lForEach{node $\in$ subnodes}{node.layout()}
\nlset{Y}y$\gets$0\;
\ForEach{node $\in$ subnodes}{%
\nlset{X}x = \Switch{alignment}{\lCase{left}{0}\lCase{centerX}{$-$node.width$/2$}\lCase{right}{$-$node.width}}
\nlset{XY}node.place(x, y)\;
\nlset{Y}y $\mathrel{+}=$ node.height $+$ spacing
}

\Return{\{\\
\nlset{X}    w: maxBy(subnodes, `width'),\\
\nlset{Y}    h: sumBy(subnodes, `height') $+$ spacing $\cdot$ $(|$subnodes$| - 1)$\\
\}}
\end{algorithm}

To make a framework with low viscosity, we want to support authoring any given graphic representation in many different ways. This property increases the malleability of the language, because a specification can be rewritten into many equivalent forms where each form may be adjacent to different specifications with new meanings.
In~\Cref{sec:tradeoff} we introduced two patterns for rewrites of this kind: denesting relations and splitting them apart. We would like a layout engine where the following three specifications are equivalent.

A purely hierarchical specification:
\begin{codeblock}
    \begin{code}
\definecolor{MineShaft}{RGB}{59,59,59}
\definecolor{Maroon}{RGB}{128,0,0}
\definecolor{JellyBean}{RGB}{38,127,153}
\definecolor{Red}{RGB}{229,0,0}
\definecolor{Black}{RGB}{0,0,0}
\definecolor{Blue}{RGB}{0,0,255}
\definecolor{Salem}{RGB}{9,134,88}
\textcolor{Maroon}{<}\textcolor{JellyBean}{StackV}\textcolor{Maroon}{>}\\
\textcolor{MineShaft}{  }\textcolor{Maroon}{<}\textcolor{JellyBean}{Rect}\textcolor{MineShaft}{ }\textcolor{Red}{width}\textcolor{Black}{=}\textcolor{Blue}{\{}\textcolor{Salem}{10}\textcolor{Blue}{\}}\textcolor{MineShaft}{ }\textcolor{Red}{height}\textcolor{Black}{=}\textcolor{Blue}{\{}\textcolor{Salem}{20}\textcolor{Blue}{\}}\textcolor{MineShaft}{ }\textcolor{Maroon}{/>}\\
\textcolor{MineShaft}{  }\textcolor{Maroon}{<}\textcolor{JellyBean}{Rect}\textcolor{MineShaft}{ }\textcolor{Red}{width}\textcolor{Black}{=}\textcolor{Blue}{\{}\textcolor{Salem}{30}\textcolor{Blue}{\}}\textcolor{MineShaft}{ }\textcolor{Red}{height}\textcolor{Black}{=}\textcolor{Blue}{\{}\textcolor{Salem}{10}\textcolor{Blue}{\}}\textcolor{MineShaft}{ }\textcolor{Maroon}{/>}\\
\textcolor{Maroon}{</}\textcolor{JellyBean}{StackV}\textcolor{Maroon}{>}
\end{code}
\end{codeblock}

\vspace{1mm}
A denested specification:
\begin{codeblock}
    \begin{code}
\definecolor{MineShaft}{RGB}{59,59,59}
\definecolor{Maroon}{RGB}{128,0,0}
\definecolor{JellyBean}{RGB}{38,127,153}
\definecolor{Red}{RGB}{229,0,0}
\definecolor{Black}{RGB}{0,0,0}
\definecolor{Tamarillo}{RGB}{163,21,21}
\definecolor{Blue}{RGB}{0,0,255}
\definecolor{Salem}{RGB}{9,134,88}
\textcolor{Maroon}{<}\textcolor{JellyBean}{Rect}\textcolor{MineShaft}{ }\textcolor{Red}{name}\textcolor{Black}{=}\textcolor{Tamarillo}{"a"}\textcolor{MineShaft}{ }\textcolor{Red}{width}\textcolor{Black}{=}\textcolor{Blue}{\{}\textcolor{Salem}{10}\textcolor{Blue}{\}}\textcolor{MineShaft}{ }\textcolor{Red}{height}\textcolor{Black}{=}\textcolor{Blue}{\{}\textcolor{Salem}{20}\textcolor{Blue}{\}}\textcolor{MineShaft}{ }\textcolor{Maroon}{/>}\\
\textcolor{Maroon}{<}\textcolor{JellyBean}{Rect}\textcolor{MineShaft}{ }\textcolor{Red}{name}\textcolor{Black}{=}\textcolor{Tamarillo}{"b"}\textcolor{MineShaft}{ }\textcolor{Red}{width}\textcolor{Black}{=}\textcolor{Blue}{\{}\textcolor{Salem}{30}\textcolor{Blue}{\}}\textcolor{MineShaft}{ }\textcolor{Red}{height}\textcolor{Black}{=}\textcolor{Blue}{\{}\textcolor{Salem}{10}\textcolor{Blue}{\}}\textcolor{MineShaft}{ }\textcolor{Maroon}{/>}\\
\textcolor{Maroon}{<}\textcolor{JellyBean}{StackV}\textcolor{Maroon}{>}\\
\textcolor{MineShaft}{  }\textcolor{Maroon}{<}\textcolor{JellyBean}{Ref}\textcolor{MineShaft}{ }\textcolor{Red}{select}\textcolor{Black}{=}\textcolor{Tamarillo}{"a"}\textcolor{MineShaft}{ }\textcolor{Maroon}{/>}\\
\textcolor{MineShaft}{  }\textcolor{Maroon}{<}\textcolor{JellyBean}{Ref}\textcolor{MineShaft}{ }\textcolor{Red}{select}\textcolor{Black}{=}\textcolor{Tamarillo}{"b"}\textcolor{MineShaft}{ }\textcolor{Maroon}{/>}\\
\textcolor{Maroon}{</}\textcolor{JellyBean}{StackV}\textcolor{Maroon}{>}
\end{code}
\end{codeblock}

A denested specification where \texttt{StackV} has been split apart:
\begin{codeblock}
    \begin{code}
\definecolor{MineShaft}{RGB}{59,59,59}
\definecolor{Maroon}{RGB}{128,0,0}
\definecolor{JellyBean}{RGB}{38,127,153}
\definecolor{Red}{RGB}{229,0,0}
\definecolor{Black}{RGB}{0,0,0}
\definecolor{Tamarillo}{RGB}{163,21,21}
\definecolor{Blue}{RGB}{0,0,255}
\definecolor{Salem}{RGB}{9,134,88}
\textcolor{Maroon}{<}\textcolor{JellyBean}{Rect}\textcolor{MineShaft}{ }\textcolor{Red}{name}\textcolor{Black}{=}\textcolor{Tamarillo}{"a"}\textcolor{MineShaft}{ }\textcolor{Red}{width}\textcolor{Black}{=}\textcolor{Blue}{\{}\textcolor{Salem}{10}\textcolor{Blue}{\}}\textcolor{MineShaft}{ }\textcolor{Red}{height}\textcolor{Black}{=}\textcolor{Blue}{\{}\textcolor{Salem}{20}\textcolor{Blue}{\}}\textcolor{MineShaft}{ }\textcolor{Maroon}{/>}\\
\textcolor{Maroon}{<}\textcolor{JellyBean}{Rect}\textcolor{MineShaft}{ }\textcolor{Red}{name}\textcolor{Black}{=}\textcolor{Tamarillo}{"b"}\textcolor{MineShaft}{ }\textcolor{Red}{width}\textcolor{Black}{=}\textcolor{Blue}{\{}\textcolor{Salem}{30}\textcolor{Blue}{\}}\textcolor{MineShaft}{ }\textcolor{Red}{height}\textcolor{Black}{=}\textcolor{Blue}{\{}\textcolor{Salem}{10}\textcolor{Blue}{\}}\textcolor{MineShaft}{ }\textcolor{Maroon}{/>}\\
\textcolor{Maroon}{<}\textcolor{JellyBean}{Distribute}\textcolor{MineShaft}{ }\textcolor{Red}{direction}\textcolor{Black}{=}\textcolor{Tamarillo}{"vertical"}\textcolor{Maroon}{>}\\
\textcolor{MineShaft}{  }\textcolor{Maroon}{<}\textcolor{JellyBean}{Ref}\textcolor{MineShaft}{ }\textcolor{Red}{select}\textcolor{Black}{=}\textcolor{Tamarillo}{"a"}\textcolor{MineShaft}{ }\textcolor{Maroon}{/>}\\
\textcolor{MineShaft}{  }\textcolor{Maroon}{<}\textcolor{JellyBean}{Ref}\textcolor{MineShaft}{ }\textcolor{Red}{select}\textcolor{Black}{=}\textcolor{Tamarillo}{"b"}\textcolor{MineShaft}{ }\textcolor{Maroon}{/>}\\
\textcolor{Maroon}{</}\textcolor{JellyBean}{Distribute}\textcolor{Maroon}{>}\\
\textcolor{Maroon}{<}\textcolor{JellyBean}{Align}\textcolor{MineShaft}{ }\textcolor{Red}{alignment}\textcolor{Black}{=}\textcolor{Tamarillo}{"centerX"}\textcolor{Maroon}{>}\\
\textcolor{MineShaft}{  }\textcolor{Maroon}{<}\textcolor{JellyBean}{Ref}\textcolor{MineShaft}{ }\textcolor{Red}{select}\textcolor{Black}{=}\textcolor{Tamarillo}{"a"}\textcolor{MineShaft}{ }\textcolor{Maroon}{/>}\\
\textcolor{MineShaft}{  }\textcolor{Maroon}{<}\textcolor{JellyBean}{Ref}\textcolor{MineShaft}{ }\textcolor{Red}{select}\textcolor{Black}{=}\textcolor{Tamarillo}{"b"}\textcolor{MineShaft}{ }\textcolor{Maroon}{/>}\\
\textcolor{Maroon}{</}\textcolor{JellyBean}{Align}\textcolor{Maroon}{>}
\end{code}
\end{codeblock}

%
We can work backwards from these equivalences to design a layout semantics that ensures these equivalences as much as possible.

\subsubsection{The \texttt{StackV} Layout Algorithm}
\Cref{alg:col}~(based on one provided by Jetpack Compose~\cite{androidCustomLayouts}), gives pseudocode for \texttt{StackV}'s layout algorithm. 
\texttt{StackV} takes as input an \texttt{alignment} (\texttt{left}, \texttt{centerX}, or \texttt{right}) and a \texttt{spacing} between elements in pixels. It then calls the \texttt{layout} algorithm of each of its children who determine their own sizes to be used later. Next, \texttt{StackV} places each of its children in its local coordinate space. The \texttt{x} coordinate refers to the left edge of the child. The \texttt{y} coordinate refers to the top edge of the child and is initialized to 0. Each child is placed horizontally based on the \texttt{alignment} parameter. In each \texttt{alignment} case, 0 is used as the guideline to which all left edges, horizontal centers, or right edges are aligned.\footnote{This guideline is in \texttt{StackV}'s local space, so the choice of 0 is arbitrary.}
%
Next, the node is placed and the next top edge is calculated by moving \texttt{spacing} pixels below the previous node. This repeats for each child. Finally, the width and height of the \texttt{StackV} are returned for use by its own parent.
%



\subsubsection{Lazy Materialization of Coordinate Transforms}
\label{sec:materialize}

\begin{figure}[]
    \centering
\includegraphics[width=0.8\linewidth]{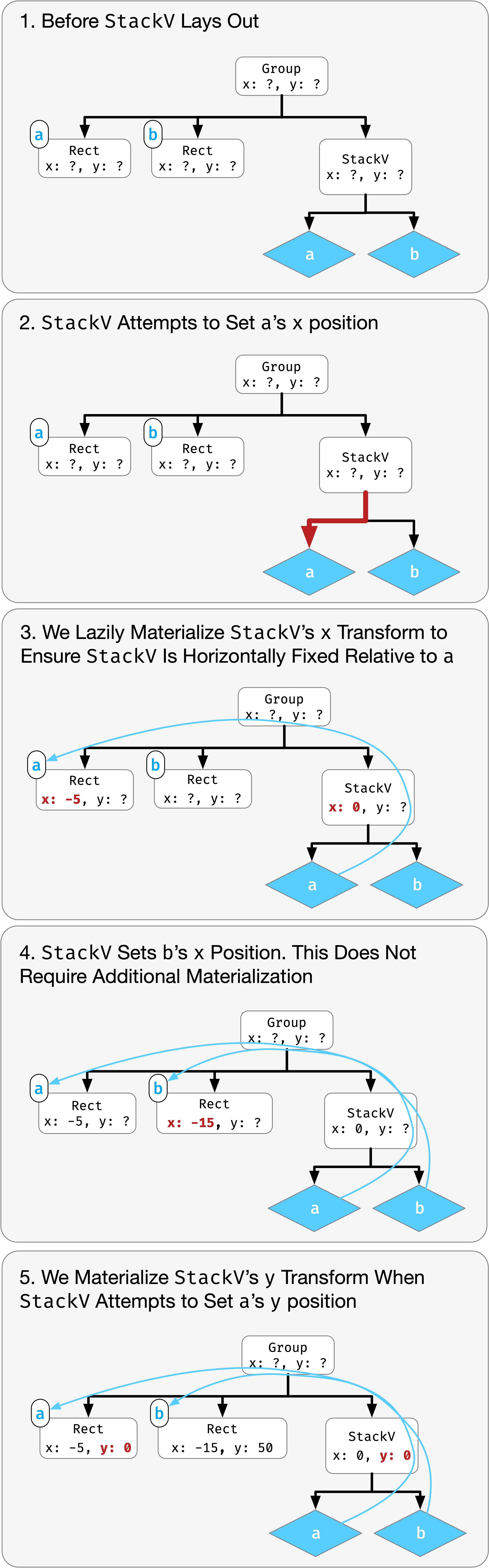}
    \caption{When a bounding box dimension is requested via a \texttt{Ref}, intermediate transforms are lazily materialized. This ensures the dimension is well-defined relative to the requester.}
    \label{fig:materialization}
    \Description{A diagram of how the lazy materialization algorithm initializes transforms in the scenegraph.
1. Before StackV lays out.
Group (x ?, y ?)
- Rect (x ?, y ?) [a]
- Rect (x ?, y ?) [b]
- StackV (x ?, y ?)
-- Ref a
-- Ref b
2. StackV attempts to set a's x position.
3. We lazily materialize StackV's x transform to ensure StackV is horizontally fixed relative to a.
Group (x ?, y ?)
- Rect (x -5, y ?)
- Rect (x ?, y ?)
- StackV (x 0, y ?)
-- Ref a
-- Ref b
4. StackV sets b's x position. This does not require additional materialization.
Group (x ?, y ?)
- Rect (x -5, y ?)
- Rect (x -15, y ?)
- StackV (x 0, y ?)
-- Ref a
-- Ref b
5. We materialize StackV's y transform when StackV attempts to set a's y position
Group (x ?, y ?)
- Rect (x -5, y 0)
- Rect (x -15, y 50)
- StackV (x 0, y 0)
-- Ref a
-- Ref b}
\end{figure}

To ensure that Bluefish is a conservative extension of UI architectures, layout algorithms like \Cref{alg:col} must work correctly even when some of their children are references, such as when denesting \texttt{StackV}.

When the \texttt{layout} method is called on a \texttt{Ref} node, it triggers reference resolution to instead return information about the bounding box pointed to by the reference.
To resolve a reference, we have to transform the bounding box of the referent into the reference's coordinate frame. We accomplish this by walking the scenegraph between the reference and the referent through their least common ancestor.
%
However, it is often the case that one or more of these intermediate nodes does not have a defined coordinate transform of its own. For example, since the \texttt{Ref} children of \texttt{StackV} are resolved
during \texttt{StackV}'s own layout algorithm.
\texttt{StackV}'s transform is not yet known.
In these cases, we must \textit{materialize} intermediate coordinate transforms. 
\Cref{fig:materialization} depicts lazy coordinate transform materialization during the layout of our running example. When \texttt{StackV} attempts to set \texttt{a}'s \texttt{x} position, its own transform is not yet known. We thus default \texttt{StackV}'s \texttt{x} transform to the identity transform. Materializing this transform ensures that the horizontal position of \texttt{StackV} is fixed relative to \texttt{a}, which helps guarantee that all of \texttt{StackV}'s children are actually vertically stacked. By deferring transform materialization lazily until a value is requested, we help ensure that the specification is as flexible as possible. If transforms were defaulted eagerly, the position of every object would be fixed before layouts could set them.


\subsubsection{Relaxing Node Ownership to Dimension Ownership}
\label{sec:ownership}

\begin{figure}[h]
    \centering
    \includegraphics[width=0.9\linewidth]{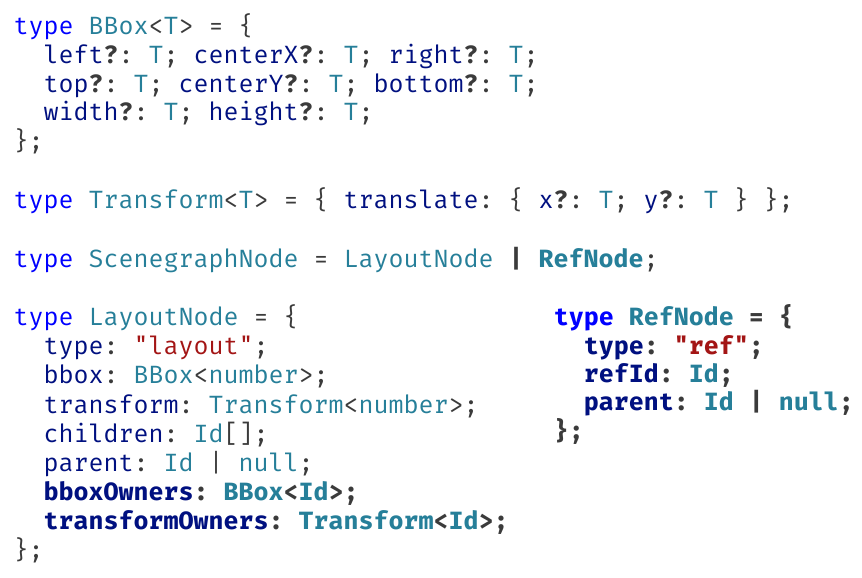}
    \caption{The TypeScript type specification for Bluefish's relational scenegraph. Bolded sections are extensions to a typical tree-structured scenegraph. Specifically, we (i) make bounding box and coordinate transform fields optional; (ii) track ownership of individual fields; and (iii) add a \texttt{RefNode} type for adjacency relations.}
    \label{fig:rsg-type}
    \Description{The TypeScript type specification of Bluefish's relational scenegraph.
type BBox<T> = {
left?: T;
centerX?: T;
right?: T;
top?: T;
centerY?: T;
bottom?: T;
width?: T;
height?: T;
};
type Transform<T> = {
translate: { x?: T; y?: T };
};
type ScenegraphNode = LayoutNode | RefNode;
type LayoutNode = {
type: "layout";
bbox: BBox<number>;
transform: Transform<number>;
children: Id[];
parent: Id | null;
bboxOwners: BBox<Id>;
transformOwners: Transform<Id>;
};
type RefNode = {
type: "ref";
refId: Id;
parent: Id | null;
};}
\end{figure}

The \texttt{StackV} layout in \Cref{alg:col} can be cleanly separated into a horizontal \texttt{Align} and a vertical \texttt{Distribute} relation by using the lines labeled \texttt{X} and \texttt{Y}, respectively, because the logic for each axis are essentially disjoint.
Though splitting \texttt{StackV}'s layout is straightforward, it creates an architectural problem. In order to split \texttt{StackV} in two, we must allow multiple relations to modify a single node.

Typically in a UI layout engine a node is owned by a single parent and only that parent's layout may modify the node. As a result, the relation established by that parent (such as a \texttt{StackV}) can never be mutated. This makes UI specifications \textit{declarative}: a relation like \texttt{StackV} always corresponds to a vertical stack in the diagram.
We want to preserve this correspondence when a node has more than one parent. To account for this multiplicity, instead of a parent node owning an entire child node, a parent node owns specific dimensions of a child node's bounding box. \Cref{fig:rsg-type} summarizes the modifications to a tree-structured scenegraph datatype required to implement bounding box ownership. Bluefish throws an error if another layout tries to write to a dimension that is already owned, which guards against overconstrained layouts (such as aligning an element to two different elements that have already been placed). Tracking the specific owner (rather than just whether or not a property is owned) allows us to determine the two layouts that conflict. Overconstrained layouts occur frequently when editing a diagram, but problems tend to be easy to localize with access to ownership information and when relations are added one at a time.

\section{Example Gallery}
\label{sec:example-gallery}

\begin{figure*}
    \centering
    \includegraphics[width=0.875\textwidth]{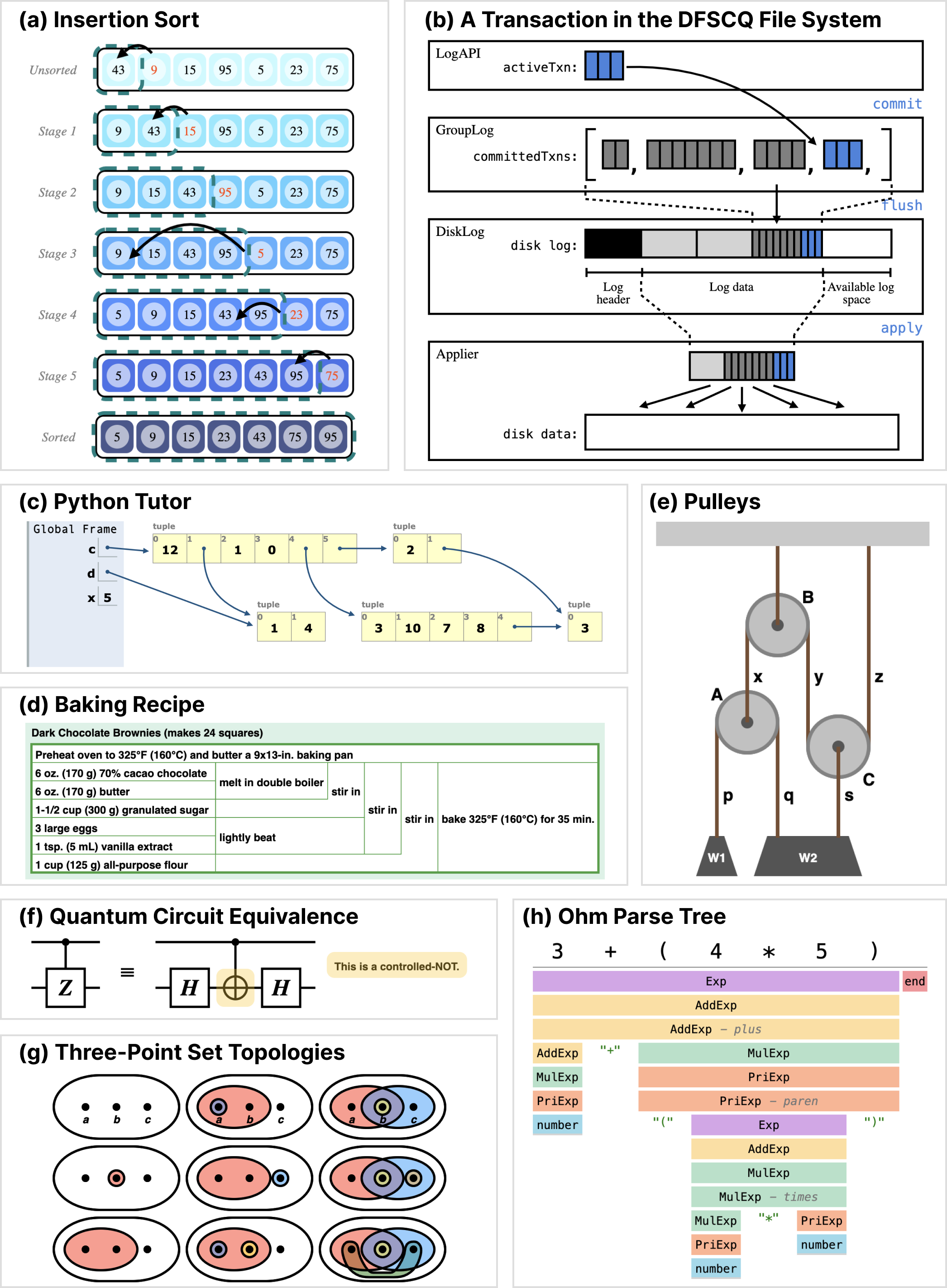}
    \caption{To evaluate Bluefish's expressiveness, we created a diverse example gallery drawn from several domains including (a--c, h) computer science, (d) cooking, (e–f) physics, and (g) math. Code and live examples are available in supplemental materials.}
    \label{fig:examples}
    \Description{Figure 8.a: A step-by-step visualization of the Insertion Sort algorithm. The diagram consists of a column of labeled arrays. Each array has a series of numbers and a bordered region representing the sorted section of the array. In each array the next item to be sorted is highlighted and an arrow points from the item to its future position in the array.

Figure 8.b: A depiction of a DFSCQ file system transaction. The transaction proceeds from LogAPI to GroupLog to DiskLog to Applier via commit, flush, and apply, respectively. Each level contains block lists arranged in different ways. Arrows connect blocks in one level to blocks in the next.

Figure 8.c: A program state diagram for a Python program. The diagram is split into a stack and a heap. The stack is a vertical array of pairs of variables and numbers or pointers into the heap. The heap is a graph of tuples, which have numbers and pointers to other tuples.

Figure 8.d: A table representing a cookie baking recipe. The leftmost column is the list of ingredients. Cells moving to the right represent steps. Each cell is sized vertically so as to contain its ingredients and preceding steps.

Figure 8.e: A pulley diagram consisting of a ceiling, pulleys, ropes, and weights. Each rope, pulley, and weight is annotated. The description given by Larkin&Simon: "The first weight is suspended from the left end of a rope over Pulley A. The right end of this rope is attached to, and partially supports, the second weight. Pulley A is suspended from the left end of a rope that runs over Pulley B, and under Pulley C. Pulley B is suspended from the ceiling. The right end of the rope that runs under Pulley C is attached to the ceiling. Pulley C is attached to the second weight, supporting it jointly with the right end of the first rope.}
\end{figure*}

To evaluate the strengths and limitations of Bluefish, we constructed a gallery of example diagrams in collaboration with a creative coder~\Cref{fig:examples}.
As there are no well-established diagram taxonomies, we instead decided to collect diagrams that are highly complex and that run the spectrum of common diagram structures including tables, overlapping containments (e.g. Venn diagrams), trees, graphs, and lists.
These examples are inspired by existing diagrams across computer science, physics, math, and cooking.
We created them using only the primitives in the Bluefish standard library with a few exceptions where we take advantage of a special \texttt{LayoutFunction} relation to sidestep current limitations of the library. Live examples and code are available in the supplemental material.

\Cref{tab:example_summary} lists the diagrams, their domains, the Gestalt relations they use, and their render times.
We first describe two examples in detail~(\Cref{sec:selected-examples}) that we use for our comparisons in \Cref{sec:comparison}. They illustrate how typical Bluefish specifications are constructed.
We then identify three general limitations of our current abstractions that we discovered when creating our gallery~(\Cref{sec:general-limitations}).
Finally, we conducted a preliminary performance evaluation by comparing Bluefish to existing baseline implementations of three diagrams from our gallery~(\Cref{sec:perf}). We find that Bluefish scales linearly with the size of its scenegraph. Bluefish is asymptotically faster than Penrose on the \textit{Insertion Sort} diagram and less than ten times slower compared to the original D3-based implementations of the \textit{Python Tutor} and \textit{Ohm Parse Tree} diagrams.
\begin{table*}[]
    \centering
    \begin{tabular}{l|c|c|c|c|c|r}
    \multirow{3}{*}{Diagram} & \multirow{3}{*}{Domain} & \multicolumn{4}{c|}{Relations} & \multirow{3}{*}{\makecell{Render\\time (ms)}} \\
    \cline{3-6}
    & & Alignment & \makecell{Uniform\\density} & Connectedness & \makecell{Common\\region} & \\\hline
       (a) Insertion Sort~\cite{penrose,githubFeatArraymanipulation} & CS & \checkmark & \checkmark & \checkmark & \checkmark & 163.56 \\
       (b) DFSCQ File System~\cite{basalt,chen2017verifying} & CS & \checkmark & \checkmark & \checkmark & \checkmark & 155.24 \\
       (c) Python Tutor~\cite{pythontutor} & CS & \checkmark & \checkmark & \checkmark & \checkmark & 149.74 \\
       (d) Baking Recipe~\cite{cookingforengineersDarkChocolate} & Cooking & \checkmark & \checkmark & - & \checkmark & 99.70 \\
       (e) Pulleys~\cite{larkin1987diagram} & Physics & \checkmark & \checkmark & \checkmark & \checkmark & 95.18 \\
       (f) Quantum Circuit~\cite{joy2020implementation} & Physics & \checkmark & \checkmark & \checkmark & \checkmark & 68.99 \\
       (g) Three-Point Set Topologies~\cite{Munkres1999-yi} & Math & \checkmark & \checkmark & - & \checkmark & 129.58 \\
       (h) Ohm Parse Tree~\cite{ohm} & CS & \checkmark & \checkmark & - & \checkmark & 174.10 \\
    \end{tabular}
    \caption{The domains, relations, marks, and render times of the diagrams in \Cref{fig:examples}. The examples demonstrate coverage over the four Gestalt relations supported by Bluefish's standard library. All examples run in less than 175ms.}
    \label{tab:example_summary}
\end{table*}

\subsection{Selected Examples}
\label{sec:selected-examples}



\subsubsection{Insertion Sort~\cite{penrose,githubFeatArraymanipulation}}

This diagram traces the steps of the insertion sort algorithm and was originally created for the Penrose~\cite{penrose} example gallery. We compare our specification to Penrose's in \Cref{sec:penrose}. The bordered region represents the sorted part of the array, and the arrow shows the insertion of the next element into the sorted region. In Bluefish we encapsulate this diagram as an element that takes an unsorted array:
\begin{codeblock}
\begin{code}
\definecolor{MineShaft}{RGB}{59,59,59}
\definecolor{Maroon}{RGB}{128,0,0}
\definecolor{JellyBean}{RGB}{38,127,153}
\definecolor{Red}{RGB}{229,0,0}
\definecolor{Black}{RGB}{0,0,0}
\definecolor{Blue}{RGB}{0,0,255}
\definecolor{Salem}{RGB}{9,134,88}
\textcolor{Maroon}{<}\textcolor{JellyBean}{InsertionSort}\\
\textcolor{MineShaft}{  }\textcolor{Red}{array}\textcolor{Black}{=}\textcolor{Blue}{\{}\textcolor{Black}{[}\textcolor{Salem}{43}\textcolor{Black}{, }\textcolor{Salem}{9}\textcolor{Black}{, }\textcolor{Salem}{15}\textcolor{Black}{, }\textcolor{Salem}{95}\textcolor{Black}{, }\textcolor{Salem}{5}\textcolor{Black}{, }\textcolor{Salem}{23}\textcolor{Black}{, }\textcolor{Salem}{75}\textcolor{Black}{]}\textcolor{Blue}{\}}\\
\textcolor{Maroon}{/>}
\end{code}
\end{codeblock}

This diagram is a good example of a deeply nested Bluefish specification built entirely with the standard library primitives.
\texttt{InsertionSortDiagram} creates a \texttt{StackV} of \texttt{InsertionSortStep}s then places labels using a series of \texttt{StackH} relations:

\begin{codeblock}
\begin{code}
\definecolor{MineShaft}{RGB}{59,59,59}
\definecolor{Blue}{RGB}{0,0,255}
\definecolor{Dallas}{RGB}{121,94,38}
\definecolor{Black}{RGB}{0,0,0}
\definecolor{NavyBlue}{RGB}{0,16,128}
\definecolor{Lochmara}{RGB}{0,112,193}
\definecolor{ElectricViolet}{RGB}{175,0,219}
\definecolor{Maroon}{RGB}{128,0,0}
\definecolor{JellyBean}{RGB}{38,127,153}
\definecolor{Red}{RGB}{229,0,0}
\definecolor{Salem}{RGB}{9,134,88}
\textcolor{Blue}{const}\textcolor{MineShaft}{ }\textcolor{Dallas}{InsertionSort}\textcolor{MineShaft}{ }\textcolor{Black}{=}\textcolor{MineShaft}{ }\textcolor{Dallas}{withBluefish}%
\textcolor{MineShaft}{(}%
\textcolor{NavyBlue}{props}%
\textcolor{MineShaft}{ }%
\textcolor{Blue}{=>}\textcolor{MineShaft}{ \{}\\
\textcolor{MineShaft}{  }\textcolor{Blue}{const}\textcolor{MineShaft}{ }\textcolor{Lochmara}{insertionSortSteps}\textcolor{MineShaft}{ }\textcolor{Black}{=}\\
\textcolor{MineShaft}{    }\textcolor{Dallas}{computeInsertionSortSteps}\textcolor{MineShaft}{(}\textcolor{NavyBlue}{props}\textcolor{MineShaft}{.}\textcolor{NavyBlue}{array}\textcolor{MineShaft}{);}\\
\textcolor{MineShaft}{  }\textcolor{ElectricViolet}{return}\textcolor{MineShaft}{ (}\\
\textcolor{MineShaft}{    }\textcolor{Maroon}{<}\textcolor{JellyBean}{Group}\textcolor{Maroon}{>}\\
\textcolor{MineShaft}{      }\textcolor{Maroon}{<}\textcolor{JellyBean}{StackV}\textcolor{MineShaft}{ }\textcolor{Red}{spacing}\textcolor{Black}{=}\textcolor{Blue}{\{}\textcolor{Salem}{15}\textcolor{Blue}{\}}\textcolor{Maroon}{>}\\
\textcolor{MineShaft}{        }\textcolor{Maroon}{<}\textcolor{JellyBean}{For}\textcolor{MineShaft}{ }\textcolor{Red}{each}\textcolor{Black}{=}\textcolor{Blue}{\{}\textcolor{Lochmara}{insertionSortSteps}\textcolor{Blue}{\}}\textcolor{Maroon}{>}\\
\textcolor{MineShaft}{          }\textcolor{Blue}{\{}\textcolor{Black}{(}\textcolor{NavyBlue}{data}\textcolor{Black}{, }\textcolor{Dallas}{i}\textcolor{Black}{) }\textcolor{Blue}{=>}\textcolor{Black}{ (}\\
\textcolor{Black}{            }\textcolor{Maroon}{<}\textcolor{JellyBean}{InsertionSortStep}\\
\textcolor{Black}{              }\textcolor{Red}{name}\textcolor{Black}{=}\textcolor{Blue}{\{}\textcolor{Dallas}{i}\textcolor{Black}{()}\textcolor{Blue}{\}}\\
\textcolor{Black}{              }\textcolor{Red}{stage}\textcolor{Black}{=}\textcolor{Blue}{\{}\textcolor{Dallas}{i}\textcolor{Black}{()}\textcolor{Blue}{\}}\\
\textcolor{Black}{              }\textcolor{Red}{data}\textcolor{Black}{=}\textcolor{Blue}{\{}\textcolor{NavyBlue}{data}\textcolor{Blue}{\}}\textcolor{Black}{ }\textcolor{Maroon}{/>}\\
\textcolor{Black}{          )}\textcolor{Blue}{\}}\\
\textcolor{MineShaft}{        }\textcolor{Maroon}{</}\textcolor{JellyBean}{For}\textcolor{Maroon}{>}\\
\textcolor{MineShaft}{      }\textcolor{Maroon}{</}\textcolor{JellyBean}{StackV}\textcolor{Maroon}{>}\\
\textcolor{MineShaft}{      }\textcolor{Maroon}{<}\textcolor{JellyBean}{For}\textcolor{MineShaft}{ }\textcolor{Red}{each}\textcolor{Black}{=}\textcolor{Blue}{\{}\textcolor{Lochmara}{insertionSortSteps}\textcolor{Blue}{\}}\textcolor{Maroon}{>}\\
\textcolor{MineShaft}{        }\textcolor{Blue}{\{}\textcolor{Black}{(}\textcolor{NavyBlue}{\_}\textcolor{Black}{, }\textcolor{Dallas}{i}\textcolor{Black}{) }\textcolor{Blue}{=>}\textcolor{Black}{ (}\\
\textcolor{Black}{          }\textcolor{Maroon}{<}\textcolor{JellyBean}{StackH}\textcolor{Black}{ }\textcolor{Red}{spacing}\textcolor{Black}{=}\textcolor{Blue}{\{}\textcolor{Salem}{20}\textcolor{Blue}{\}}\textcolor{Maroon}{>}\\
\textcolor{Black}{            }\textcolor{Maroon}{<}\textcolor{JellyBean}{LabelText}\textcolor{Maroon}{>}\\
\textcolor{Black}{              }\textcolor{Blue}{\{}\textcolor{Dallas}{label}\textcolor{Black}{(}\textcolor{Dallas}{i}\textcolor{Black}{(), }\textcolor{NavyBlue}{props}\textcolor{Black}{.}\textcolor{NavyBlue}{array}\textcolor{Black}{.}\textcolor{NavyBlue}{length}\textcolor{Black}{)}\textcolor{Blue}{\}}\\
\textcolor{Black}{            }\textcolor{Maroon}{</}\textcolor{JellyBean}{LabelText}\textcolor{Maroon}{>}\\
\textcolor{Black}{            }\textcolor{Maroon}{<}\textcolor{JellyBean}{Ref}\textcolor{Black}{ }\textcolor{Red}{select}\textcolor{Black}{=}\textcolor{Blue}{\{}\textcolor{Dallas}{i}\textcolor{Black}{()}\textcolor{Blue}{\}}\textcolor{Black}{ }\textcolor{Maroon}{/>}\\
\textcolor{Black}{          }\textcolor{Maroon}{</}\textcolor{JellyBean}{StackH}\textcolor{Maroon}{>}\\
\textcolor{Black}{        )}\textcolor{Blue}{\}}\\
\textcolor{MineShaft}{      }\textcolor{Maroon}{</}\textcolor{JellyBean}{For}\textcolor{Maroon}{>}\\
\textcolor{MineShaft}{    }\textcolor{Maroon}{</}\textcolor{JellyBean}{Group}\textcolor{Maroon}{>}\\
\textcolor{MineShaft}{  )\})}
\end{code}
\end{codeblock}

Like \texttt{InsertionSort}, \texttt{InsertionSortStep} is a custom element. It specifies a \texttt{StackH} of custom \texttt{ArrayEntry} elements inside a \texttt{Background}, then uses a custom \texttt{DashedBorder} relation that specializes \texttt{Background} to surround the sorted region of the array, and finally uses an \texttt{Arrow} relation to show the movement of each element into the sorted region.




\subsubsection{DFSCQ File System~\cite{basalt,chen2017verifying}}
This diagram describes the life of a transaction in the DFSCQ file system. The diagram was originally created in Inkscape and recreated in the Basalt diagramming framework to test the limits of its expressiveness~\cite{basalt}. We compare our specification to Basalt's in \Cref{sec:basalt}. As with the \textit{Insertion Sort} diagram, the \textit{DFSCQ File System} diagram's specification uses several custom elements and relations composed of Bluefish standard library primitives. For example, the top level specification consists of a \texttt{StackV} containing four custom \texttt{TitledBackground} relations interspersed with custom \texttt{ActionText} elements like these:

\begin{codeblock}
\begin{code}
\definecolor{MineShaft}{RGB}{59,59,59}
\definecolor{Maroon}{RGB}{128,0,0}
\definecolor{JellyBean}{RGB}{38,127,153}
\definecolor{Red}{RGB}{229,0,0}
\definecolor{Black}{RGB}{0,0,0}
\definecolor{Tamarillo}{RGB}{163,21,21}
\definecolor{Blue}{RGB}{0,0,255}
\definecolor{Salem}{RGB}{9,134,88}
\textcolor{Maroon}{<}\textcolor{JellyBean}{TitledBackground}\textcolor{MineShaft}{ }\textcolor{Red}{title}\textcolor{Black}{=}\textcolor{Tamarillo}{"LogAPI"}\textcolor{Maroon}{>}\\
\textcolor{MineShaft}{  }\textcolor{Maroon}{<}\textcolor{JellyBean}{StackH}\textcolor{Maroon}{>}\\
\textcolor{MineShaft}{    }\textcolor{Maroon}{<}\textcolor{JellyBean}{BoxedAlign}\textcolor{MineShaft}{ }\textcolor{Red}{alignment}\textcolor{Black}{=}\textcolor{Tamarillo}{"centerRight"}\\
\textcolor{MineShaft}{               }\textcolor{Red}{width}\textcolor{Black}{=}\textcolor{Blue}{\{}\textcolor{Salem}{200}\textcolor{Blue}{\}}\textcolor{Maroon}{>}\\
\textcolor{MineShaft}{      }\textcolor{Maroon}{<}\textcolor{JellyBean}{Text}\textcolor{MineShaft}{ }\textcolor{Red}{font-family}\textcolor{Black}{=}\textcolor{Tamarillo}{"monospace"}\\
\textcolor{MineShaft}{            }\textcolor{Red}{font-weight}\textcolor{Black}{=}\textcolor{Blue}{\{}\textcolor{Salem}{300}\textcolor{Blue}{\}}\\
\textcolor{MineShaft}{            }\textcolor{Red}{font-size}\textcolor{Black}{=}\textcolor{Tamarillo}{"18"}\textcolor{Maroon}{>}\\
\textcolor{MineShaft}{        activeTxn:}\\
\textcolor{MineShaft}{      }\textcolor{Maroon}{</}\textcolor{JellyBean}{Text}\textcolor{Maroon}{>}\\
\textcolor{MineShaft}{    }\textcolor{Maroon}{</}\textcolor{JellyBean}{BoxedAlign}\textcolor{Maroon}{>}\\
\textcolor{MineShaft}{    }\textcolor{Maroon}{<}\textcolor{JellyBean}{Blocks}\\
\textcolor{MineShaft}{      }\textcolor{Red}{colors}\textcolor{Black}{=}\textcolor{Blue}{\{}\textcolor{Black}{[}\textcolor{Tamarillo}{"\#4582DE"}\textcolor{Black}{, }\textcolor{Tamarillo}{"\#4582DE"}\textcolor{Black}{, }\textcolor{Tamarillo}{"\#4582DE"}\textcolor{Black}{]}\textcolor{Blue}{\}}\\
\textcolor{MineShaft}{      }\textcolor{Red}{name}\textcolor{Black}{=}\textcolor{Tamarillo}{"activeTxnBlock"}\textcolor{MineShaft}{ }\textcolor{Maroon}{/>}\\
\textcolor{MineShaft}{  }\textcolor{Maroon}{</}\textcolor{JellyBean}{StackH}\textcolor{Maroon}{>}\\
\textcolor{Maroon}{</}\textcolor{JellyBean}{TitledBackground}\textcolor{Maroon}{>}\\
\textcolor{Maroon}{<}\textcolor{JellyBean}{ActionText}\textcolor{MineShaft}{ }\textcolor{Red}{text}\textcolor{Black}{=}\textcolor{Tamarillo}{"commit"}\textcolor{MineShaft}{ }\textcolor{Maroon}{/>}
\end{code}
\end{codeblock}

\texttt{TitledBackground} is a custom relation composed of Bluefish standard library primitives:

\begin{codeblock}
\begin{code}
\definecolor{MineShaft}{RGB}{59,59,59}
\definecolor{Blue}{RGB}{0,0,255}
\definecolor{Dallas}{RGB}{121,94,38}
\definecolor{Black}{RGB}{0,0,0}
\definecolor{NavyBlue}{RGB}{0,16,128}
\definecolor{Maroon}{RGB}{128,0,0}
\definecolor{JellyBean}{RGB}{38,127,153}
\definecolor{Red}{RGB}{229,0,0}
\definecolor{Tamarillo}{RGB}{163,21,21}
\definecolor{Salem}{RGB}{9,134,88}
\textcolor{Blue}{const}\textcolor{MineShaft}{ }\textcolor{Dallas}{TitledBackground}\textcolor{MineShaft}{ }\textcolor{Black}{=}\textcolor{MineShaft}{ }\textcolor{Dallas}{withBluefish}\textcolor{MineShaft}{(}\textcolor{NavyBlue}{props}\textcolor{MineShaft}{ }\textcolor{Blue}{=>}\textcolor{MineShaft}{ (}\\
\textcolor{MineShaft}{  }\textcolor{Maroon}{<}\textcolor{JellyBean}{Align}\textcolor{MineShaft}{ }\textcolor{Red}{alignment}\textcolor{Black}{=}\textcolor{Tamarillo}{"topLeft"}\textcolor{Maroon}{>}\\
\textcolor{MineShaft}{    }\textcolor{Maroon}{<}\textcolor{JellyBean}{Text}\textcolor{MineShaft}{ }\textcolor{Red}{font-family}\textcolor{Black}{=}\textcolor{Tamarillo}{"serif" }\textcolor{Red}{font-weight}\textcolor{Black}{=}\textcolor{Blue}{\{}\textcolor{Salem}{300}\textcolor{Blue}{\}}\\
\textcolor{MineShaft}{          }\textcolor{Red}{font-size}\textcolor{Black}{=}\textcolor{Tamarillo}{"20" }\textcolor{Red}{x}\textcolor{Black}{=}\textcolor{Blue}{\{}\textcolor{Salem}{10}\textcolor{Blue}{\}}\textcolor{MineShaft}{ }\textcolor{Red}{y}\textcolor{Black}{=}\textcolor{Blue}{\{}\textcolor{Salem}{4}\textcolor{Blue}{\}}\textcolor{Maroon}{>}\\
\textcolor{MineShaft}{      }\textcolor{Blue}{\{}\textcolor{NavyBlue}{props}\textcolor{Black}{.}\textcolor{NavyBlue}{title}\textcolor{Blue}{\}}\\
\textcolor{MineShaft}{    }\textcolor{Maroon}{</}\textcolor{JellyBean}{Text}\textcolor{Maroon}{>}\\
\textcolor{MineShaft}{    }\textcolor{Maroon}{<}\textcolor{JellyBean}{Background}\textcolor{MineShaft}{ }\textcolor{Red}{padding}\textcolor{Black}{=}\textcolor{Blue}{\{}\textcolor{Salem}{30}\textcolor{Blue}{\}}\textcolor{Maroon}{>}\\
\textcolor{MineShaft}{      }\textcolor{Maroon}{<}\textcolor{JellyBean}{Align}\textcolor{MineShaft}{ }\textcolor{Red}{alignment}\textcolor{Black}{=}\textcolor{Tamarillo}{"centerLeft"}\textcolor{Maroon}{>}\\
\textcolor{MineShaft}{        }\textcolor{Maroon}{<}\textcolor{JellyBean}{Rect}\textcolor{MineShaft}{ }\textcolor{Red}{height}\textcolor{Black}{=}\textcolor{Blue}{\{}\textcolor{Salem}{0}\textcolor{Blue}{\}}\textcolor{MineShaft}{ }\textcolor{Red}{width}\textcolor{Black}{=}\textcolor{Blue}{\{}\textcolor{Salem}{680}\textcolor{Blue}{\}}\\
\textcolor{MineShaft}{              }\textcolor{Red}{fill}\textcolor{Black}{=}\textcolor{Tamarillo}{"transparent"}\textcolor{MineShaft}{ }\textcolor{Maroon}{/>}\\
\textcolor{MineShaft}{        }\textcolor{Blue}{\{}\textcolor{NavyBlue}{props}\textcolor{Black}{.}\textcolor{NavyBlue}{children}\textcolor{Blue}{\}}\\
\textcolor{MineShaft}{      }\textcolor{Maroon}{</}\textcolor{JellyBean}{Align}\textcolor{Maroon}{>}\\
\textcolor{MineShaft}{    }\textcolor{Maroon}{</}\textcolor{JellyBean}{Background}\textcolor{Maroon}{>}\\
\textcolor{MineShaft}{  }\textcolor{Maroon}{</}\textcolor{JellyBean}{Align}\textcolor{Maroon}{>}\\
\textcolor{MineShaft}{))}
\end{code}
\end{codeblock}

\begin{figure*}
    \centering
    \begin{subfigure}[]{0.33\linewidth}
         \centering
         \includegraphics[width=\textwidth]{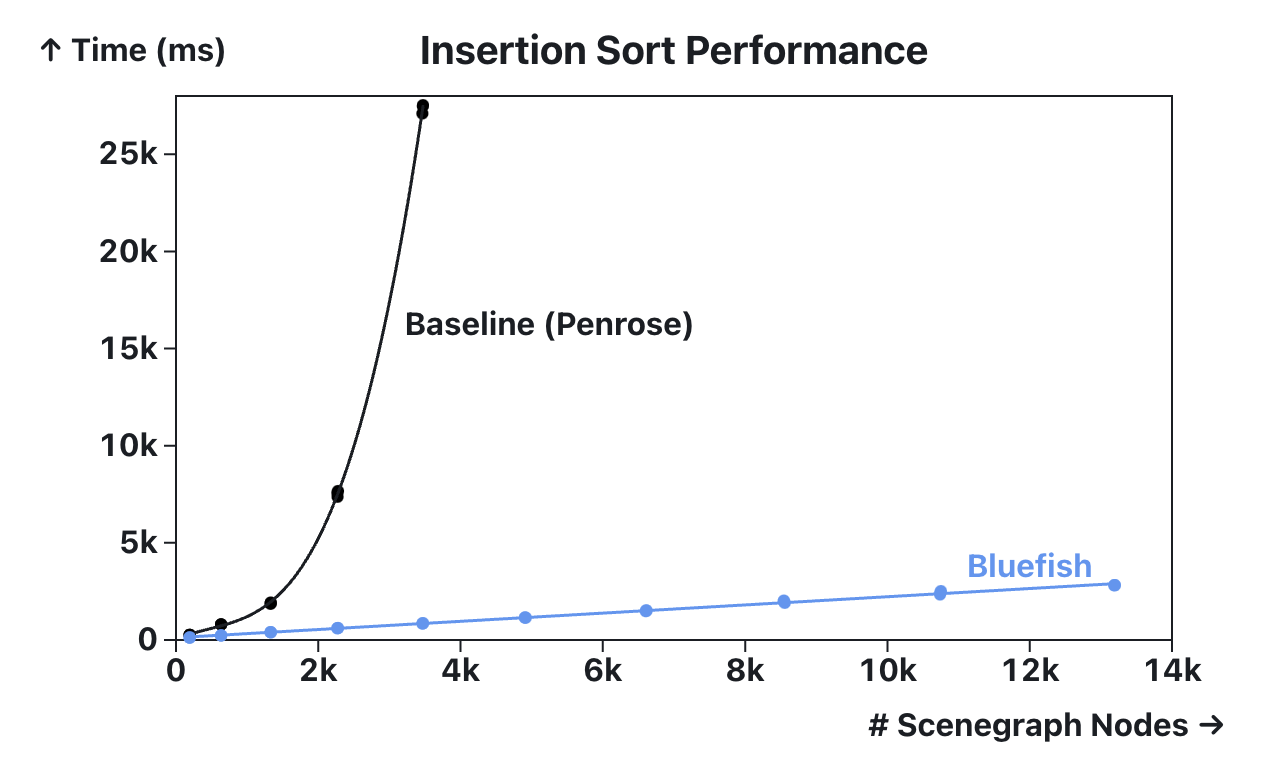}
     \end{subfigure}
    \begin{subfigure}[]{0.33\linewidth}
         \centering
         \includegraphics[width=\textwidth]{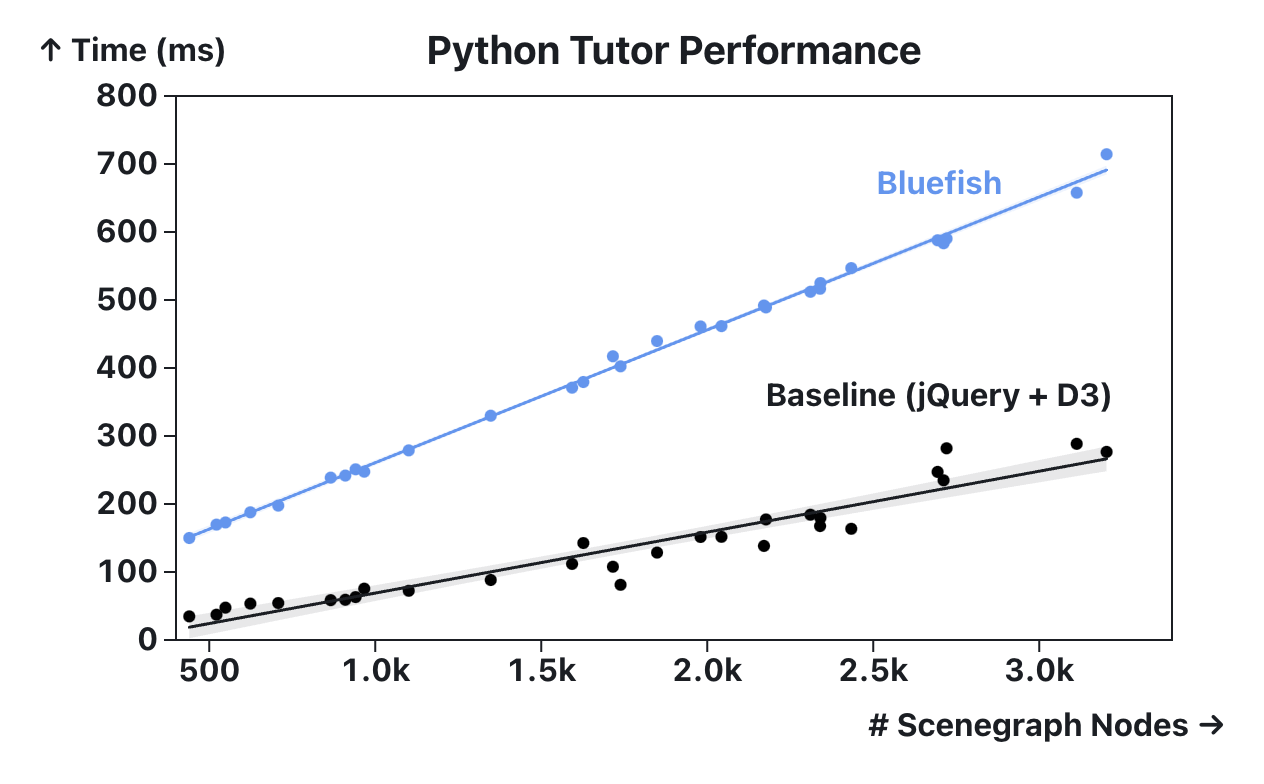}
     \end{subfigure}
    \begin{subfigure}[]{0.33\linewidth}
         \centering
         \includegraphics[width=\textwidth]{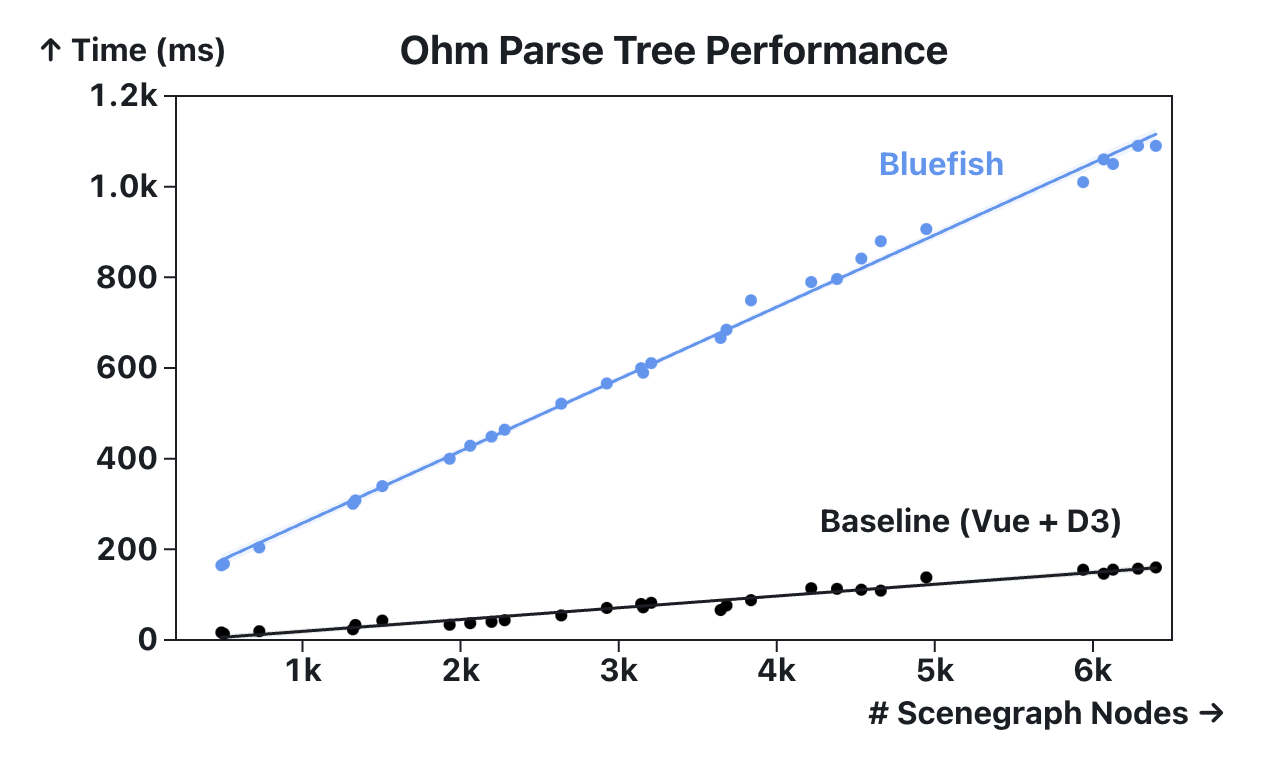}
     \end{subfigure}
    \caption{We evaluated Bluefish's performance on three examples against their original implementations. On \textit{Insertion Sort}, Bluefish scales linearly with its scenegraph size whereas Penrose scales superlinearly. Bluefish is roughly 2x and 6x slower than the original D3-based implementations of the \textit{Python Tutor} and \textit{Ohm Parse Tree} diagrams, respectively.}
    \label{fig:perf}
    \Description{Figure 9.a: This graph plots the Insertion Sort diagram. Execution Time in milliseconds is plotted on the y-axis ranging from 0 to 25,000 ms. The Number of Scenegraph Nodes is plotted on the x-axis ranging from 0 to 14,000 nodes. Bluefish scales linearly with its scenegraph size, never rising above 5,000 ms whereas Penrose scales superlinearly and reaches over 27,000 ms at just 3,000 nodes.

Figure 9.b: This graph plots the Python Tutor diagram. Execution Time in milliseconds is plotted on the y-axis ranging from 0 to 800 ms. The Number of Scenegraph Nodes is plotted on the x-axis ranging from 0 to 3,000 nodes. Bluefish scales linearly with its scenegraph size about two times slower than the baseline implementation, running in 700 ms at just over 3,000 nodes.

Figure 9.c: This graph plots the Ohm Parse Tree diagram. Execution Time in milliseconds is plotted on the y-axis ranging from 0 to 1,200 ms. The Number of Scenegraph Nodes is plotted on the x-axis ranging from 0 to 6,000 nodes. Bluefish scales linearly with its scenegraph size about six times slower than the baseline implementation, running in about 1100 ms at just over 6,000 nodes.}
\end{figure*}

After these have been placed, we place the various lines and labels that cut across this hierarchy including the arrows connecting neighboring \texttt{TitledBackground}s. To create the fanned arrows, we create a \texttt{StackH} of invisible \texttt{Rect} marks to represent different regions of the disk and connect each arrow to a different region. We ran into limitations with Bluefish for aligning the widths of the backgrounds across the four rows~(\Cref{sec:general-limitations}). To address this limitation, we used a special \texttt{LayoutFunction} relation for more expressive bounding box constraints.

\subsection{General Limitations}
\label{sec:general-limitations}
We encountered three recurring limitations while making our gallery.

\textbf{Width and Height Alignment.} 
We needed to align widths and heights of elements in several of our diagrams. For example, in the \textit{Baking Recipe} diagram, the widths of all the backgrounds in the ingredients column must be the same size, and similarly for the backgrounds in the \textit{DFSCQ File System} diagram.
At first, aligning sizes may appear to be a straightforward extension of the \texttt{Align} relation.
But unlike vertical and horizontal positions, which can often be any value, the width and height of an element typically must be large enough to contain their child elements. For example, the \texttt{Background} behind each ingredient should not be smaller than the text it contains. Aligning the size of multiple elements therefore requires determining their minimum sizes before performing layout.
%
%
UI frameworks circumvent this problem by allowing a parent to query its children's preferred sizes before performing layout. 
We could adopt a similar approach.


\textbf{Precise Alignment and Spacing.}
We ran into limitations when specifying more precise element positioning. For example, the Bluefish standard library does not including a \texttt{Padding} relation, so we sometimes used a \texttt{Stack} with an invisible \texttt{Rect} to shift elements. This workaround could be encapsulated in a custom relation.
%
Similarly, \texttt{Align}'s vertical alignments\dash{}\texttt{top}, \texttt{centerY}, and \texttt{bottom}\dash{}are not sufficient for more precise alignments with text. One often wants to align to text's visual baseline rather than the bottom of its bounding box. We worked around this problem by manually nudging text slightly in several examples. UI frameworks provide extensible guideline abstractions for working with text and images. We could adapt these solutions to Bluefish, and we have already prototyped this feature.

\textbf{Boundary Curve Abstraction.}
We also ran into limitations of Bluefish's bounding box shape abstraction. For example, when labelling the pulley in the \textit{Pulleys} diagram and the points in \textit{Three-Point Set Topologies}, we manually adjusted the text to avoid intersecting other shapes. Similarly, we manually constructed the concave set in the bottom right of the \textit{Three-Point Set Topologies} diagram to avoid overlapping the purple region.
These nudges cannot be done automatically, because Bluefish represents shapes during layout using axis-aligned bounding boxes, which are too coarse for shapes like circles or paths.
These examples suggest the need for a boundary curve abstraction. Such an abstraction would allow a user to specify that two shapes should be nested or be made disjoint with greater precision than our bounding box model. It would also support precise labeling along curved lines and arrows. We have made experimental extensions to the system that probe this idea, and we believe this is a promising future approach.

\subsection{Performance}
\label{sec:perf}

Though performance was not the primary focus of Bluefish's design,
we conducted a preliminary evaluation to assess the potential impact of Bluefish's expressiveness on performance.
Every example in our gallery renders in under 175ms. Because Bluefish executes each layout node once, we hypothesized that Bluefish's performance scales linearly with the number of scenegraph nodes. To test this, we ran the \textit{Insertion Sort}, \textit{Python Tutor}, and \textit{Ohm Parse Tree} diagrams with different input data since they have existing data-driven baseline implementations. The \textit{Insertion Sort} diagram was originally written in Penrose. The \textit{Python Tutor} and \textit{Ohm Parse Tree} diagrams were both originally written in a UI framework using D3. We evaluated these diagrams on an Apple M1 Pro SoC with 32GB of RAM using Chrome Version 126. We used the console's performance analysis
to measure the total time required to layout and render each diagram.


\Cref{fig:perf} visualizes the results of our performance testing. We found that the render time for all three diagrams scaled linearly with the number of nodes in the scenegraph. Compared to Penrose on \textit{Insertion Sort}, Bluefish scales linearly while Penrose scales superlinearly. Bluefish is roughly 2x slower than the original \textit{Python Tutor} implementation and roughly 6x slower than the original \textit{Ohm Parse Tree} implementation. These results suggest the expressiveness of Bluefish's relation abstraction preserves the linear performance scaling of local propagation and UI layout architectures. Future work can improve the constant factor overhead and investigate incremental layout performance to facilitate real-time interaction and animation use cases.



\section{Comparison to Other Compositional Approaches}
\label{sec:comparison}

In this section, we use our selected examples from \Cref{sec:selected-examples} to compare Bluefish to two recent diagramming frameworks researchers have developed. This complements the comparison to UI frameworks we conducted in \Cref{sec:components}.

\subsection{Penrose: Substance + Style}
\label{sec:penrose}

Penrose is a programming language for creating mathematical diagrams~\cite{penrose}. It features three languages: \textit{Substance} for specifying the content of a diagram, \textit{Domain} for defining the content primitives, and \textit{Style} for visualizing the Substance specification.

\begin{figure}
    \centering
    \includegraphics[width=0.45 \textwidth]{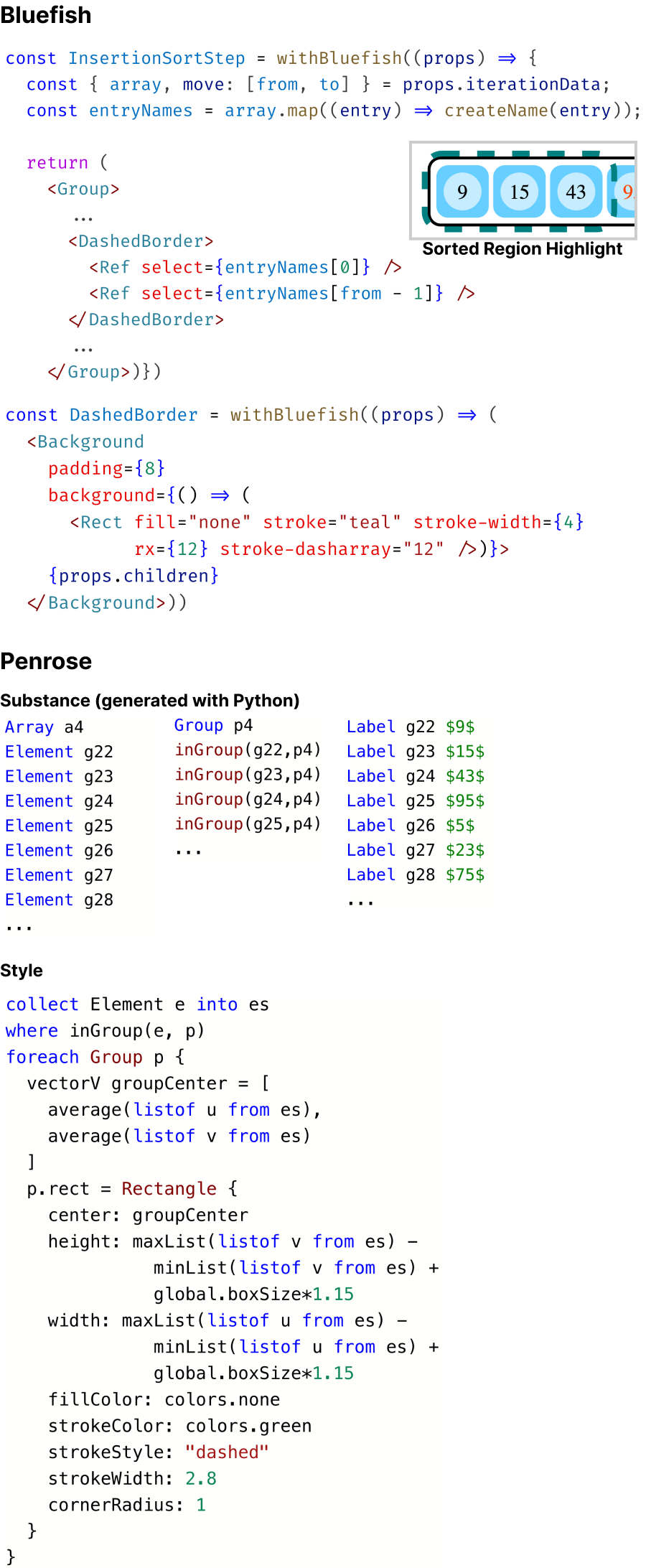}
    \caption{A comparison between Bluefish and Penrose's specifications of the dashed border in the \textit{Insertion Sort} diagram. Bluefish uses a declarative, component-based approach while Penrose's Style language draws inspiration from CSS and uses low-level bounding box calculations.}
    \label{fig:penrosevbluefish}
    \Description{A comparison of Bluefish and Penrose snippets for specifying the border around the sorted region of the array in the Insertion Sort diagram.
Bluefish code:
const InsertionSortStep = withBluefish((props) => {
const { array, move: [from, to] } = props.iterationData;
const entryNames = array.map((entry) => createName(entry));
return (
<Group>
...
<DashedBorder>
<Ref select={entryNames[0]} />
<Ref select={entryNames[from - 1]} />
</DashedBorder>
...
</Group>
)
})
const DashedBorder = withBluefish((props) => (
<Background
padding={8}
background={() => (
<Rect fill="none" stroke="teal" stroke-width={4}
rx={12} stroke-dasharray="12" />
)}
{props.children}
</Background>
))
Penrose code:
Substance (generated with Python):
Array a4
Group p4
Label g22 $9$
Element g22
inGroup(g22,p4)
Label g23 $15$
Element g23
inGroup(g23,p4)
Label g24 $43$
Element g24
inGroup(g24,p4)
Label g25 $95$
Element g25
inGroup(g25,p4)
Label g26 $5$
Element g26
...
Element g27
Label g27 $23$
Element g28
Label g28 $75$
...
Style:
collect Element e into es
where inGroup(e, p)
foreach Group p {
vectorV groupCenter = [
average(listof u from es),
average(listof v from es)
]
p.rect = Rectangle {
center: groupCenter
height: maxList(listof v from es) -
minList(listof v from es) +
global.boxSize*1.15
width: maxList(listof u from es) -
minList(listof u from es) +
global.boxSize*1.15
fillColor: colors.none
strokeColor: colors.green
strokeStyle: "dashed"
strokeWidth: 2.8
cornerRadius: 1
}
}}
\end{figure}







\subsubsection{Language Design}

\Cref{fig:penrosevbluefish} compares a snippet of the Bluefish and Penrose specifications of the sorted region highlight in the \textit{Insertion Sort} diagram. The Bluefish code uses a custom \texttt{DashedBorder} relation that encapsulates a customized \texttt{Background} relation. This relation is then used to contain the first and last entry of the sorted region. The Penrose Substance file (generated automatically from a Python script) establishes the elements and relations visualized in the diagram. These include an \texttt{Array}, the array's \texttt{Element}s, a \texttt{Group} of the sorted elements, the \texttt{inGroup} relation between elements and the group, and \texttt{Label}s for the array elements. The Style program \textit{selects} the elements of the group, collecting them into a variable \texttt{es}, and constructs a \texttt{Rectangle} that contains them.

The specifications differ primarily in how the code is organized. Bluefish colocates related data (via props) and display logic. Penrose colocates all of the data and all of the display logic in Substance and Style files, respectively. By colocating data and logic, Bluefish allows a user to encapsulate reusable pieces as custom relations like \texttt{DashedBorder}. Furthermore, Bluefish encapsulates low-level bounding box calculations behind primitive relations. By separating data and display logic, Penrose allows a user to more easily restyle an entire diagram. For mathematical domains like Euclidean geometry, which have a fixed set of primitive elements and relations, this separation is especially useful. It also frees the Substance language from conforming to a component-based syntax, which allows it to more easily match math notation.



The differences between Penrose and Bluefish stem directly from the inspirations for each system. Penrose's Substance and Style languages are loosely inspired by HTML and CSS, which similarly separate content and display logic into two DSLs. Meanwhile, Bluefish is inspired by UI component frameworks like React, which are specifically designed to couple related HTML, CSS, and JS together~\cite{reactdocseparationofconcerns} as well as take advantage of the expressiveness of a general-purpose host language~\cite{reacttalk}.

\subsubsection{Layout Engine}

Penrose's layout engine uses L-BFGS, a global solver. The \textit{Insertion Sort} diagram, while deeply nested, does not contain constraints that show the full power of Penrose. This engine can easily encode constraints that are useful for geometry like ensuring the angles of a triangle are at least 30 degrees, that labels do not overlap, or that arbitrary shapes are contained inside a circle.

As a result of its more powerful layout engine, Penrose can express more perceptual relations than Bluefish including geometric relations like line-line intersection.
Bluefish's standard library and internal node abstractions would have to be significantly extended to support these relations. Even then, it would not be easy to incorporate minimum angle requirements into a local propagation solver, because they often must be solved globally. 
Future work may investigate whether how to integrate global solvers as sublanguages within Bluefish to extend its relational expressiveness.

\subsection{Basalt: Components + Constraints}
\label{sec:basalt}

Basalt is a diagramming DSL embedded in Python~\cite{basalt}. It is a modern exemplar of languages that extend a component model with a flexible constraint system~\cite{myers1995garnet,myers1991graphical,barth1986objectGROW,henry1988usingAPOGEE}. Basalt authors create Python classes similar to UI framework components. However, they can also author \textit{constraints} to relate information between components.

\begin{figure}
    \centering
    \includegraphics[width=0.45 \textwidth]{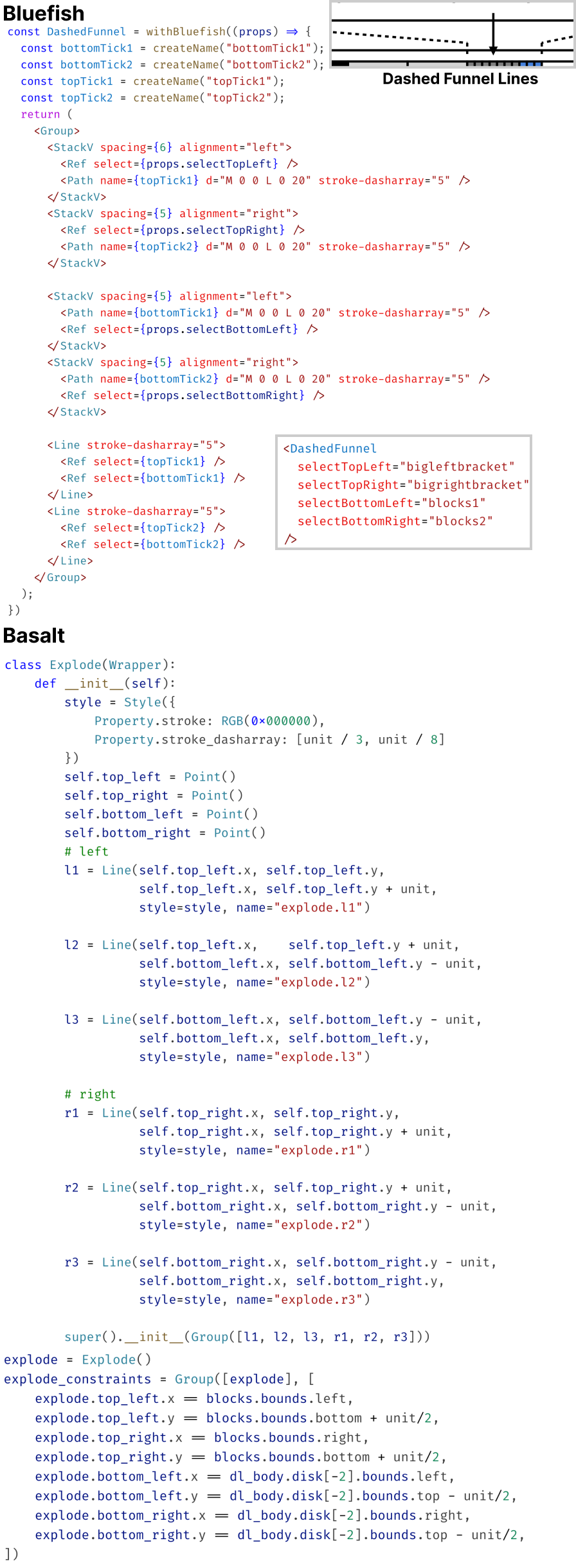}
    \caption{A comparison between Bluefish and Basalt's specifications of the dashed funnel lines connecting neighboring rows in the \textit{DFSCQ} diagram. Bluefish uses a more declarative abstraction while Basalt uses low-level constraints.}
    \label{fig:basaltvbluefish}
    \Description{A comparison of Bluefish and Basalt code for specifying the dashed funnel arrows connecting two of the levels in the DFSCQ diagram.
Bluefish code:
const DashedFunnel = withBluefish((props) => {
const bottomTick1 = createName("bottomTick1");
const bottomTick2 = createName("bottomTick2");
const topTick1 = createName("topTick1");
const topTick2 = createName("topTick2");
return (
<Group>
<StackV spacing={6} alignment="left">
<Ref select={props.selectTopLeft} />
<Path name={topTick1} d="M 0 0 L 0 20" stroke-dasharray="5" />
</StackV>
<StackV spacing={5} alignment="right">
<Ref select={props.selectTopRight} />
<Path name={topTick2} d="M 0 0 L 0 20" stroke-dasharray="5" />
</StackV>
<StackV spacing={5} alignment="left">
<Path name={bottomTick1} d="M 0 0 L 0 20" stroke-dasharray="5" />
<Ref select={props.selectBottomLeft} />
</StackV>
<StackV spacing={5} alignment="right">
<Path name={bottomTick2} d="M 0 0 L 0 20" stroke-dasharray="5" />
<Ref select={props.selectBottomRight} />
</StackV>
<Line stroke-dasharray="5">
<Ref select={topTick1} />
<Ref select={bottomTick1} />
</Line>
<Line stroke-dasharray="5">
<Ref select={topTick2} />
<Ref select={bottomTick2} />
</Line>
</Group>
);
});
<DashedFunnel
selectTopLeft="bigleftbracket"
selectTopRight="bigrightbracket"
selectBottomLeft="blocks1"
selectBottomRight="blocks2"
/>
Basalt code:
class Explode(Wrapper):
def __init__(self):
style = Style({
Property.stroke: RGB(0x000000),
Property.stroke_dasharray: [unit / 3, unit / 8]
})
self.top_left = Point()
self.top_right = Point()
self.bottom_left = Point()
self.bottom_right = Point()
# left
l1 = Line(self.top_left.x, self.top_left.y,
self.top_left.x, self.top_left.y + unit,
style=style, name="explode.l1")
l2 = Line(self.top_left.x, self.top_left.y + unit,
self.bottom_left.x, self.bottom_left.y - unit,
style=style, name="explode.l2")
l3 = Line(self.bottom_left.x, self.bottom_left.y - unit,
self.bottom_left.x, self.bottom_left.y,
style=style, name="explode.l3")
# right
r1 = Line(self.top_right.x, self.top_right.y,
self.top_right.x, self.top_right.y + unit,
style=style, name="explode.r1")
r2 = Line(self.top_right.x, self.top_right.y + unit,
self.bottom_right.x, self.bottom_right.y - unit,
style=style, name="explode.r2")
r3 = Line(self.bottom_right.x, self.bottom_right.y - unit,
self.bottom_right.x, self.bottom_right.y,
style=style, name="explode.r3")
super().__init__(Group([l1, l2, l3, r1, r2, r3]))
explode = Explode()
explode_constraints = Group([explode], [
explode.top_left.x == blocks.bounds.left,
explode.top_left.y == blocks.bounds.bottom + unit/2,
explode.top_right.x == blocks.bounds.right,
explode.top_right.y == blocks.bounds.bottom + unit/2,
explode.bottom_left.x == d1_body.disk[-2].bounds.left,
explode.bottom_left.y == d1_body.disk[-2].bounds.top - unit/2,
explode.bottom_right.x == d1_body.disk[-2].bounds.right,
explode.bottom_right.y == d1_body.disk[-2].bounds.top - unit/2,
])}
\end{figure}

\subsubsection{Language Design}

\Cref{fig:basaltvbluefish} compares a snippet of the Bluefish and Basalt specifications for the dashed funnel linkages in the \textit{DFSCQ} diagram. The Bluefish relation takes as input four names that are used to relatively position the funnel. The Basalt code instead sets up four abstract \texttt{Point}s inside the \texttt{Explode} component and aligns them to other components using constraints defined outside the component.

The Bluefish specification is more declarative. Rather than using bounding box and point constraints, the Bluefish code uses \texttt{StackV}s to position the funnel.
The Basalt specification, while lower level, is more malleable. The \texttt{Explode} component merely defines \texttt{Point}s that represent the corners of the bounding box and \texttt{Line}s related to those  corners. It leaves the positioning of the corner \texttt{Point}s for later. While the constraints could be moved inside the \texttt{Explode} component to look more like the Bluefish specification, the Bluefish specification must explicitly take in the corners as dependencies. To allow Bluefish to move the corner dependencies outside, we would have to introduce a way to position elements relative to their enclosing container and have those relations run only after the container's size has been set.

This example highlights a viscosity tradeoff between the two systems. In Bluefish, authoring specifications is more high-level as a user can think in terms of relations like \texttt{Stack} and \texttt{Line}. However, extending Bluefish with new kinds of primitive relations often requires stepping down to the low-level layout API. On the other hand, systems like Basalt are more viscous for end-users, because they must deal with constraints. But creating custom constraints is much more straightforward, because the user is already working with an expressive, low-level API. Future work may explore whether the locality-expressiveness tradeoffs we highlighted in \Cref{sec:tradeoff} could be extended to the level of bounding box and point constraints.






\subsubsection{Layout Engine}

Basalt uses Z3~\cite{de2008z3}, an SMT solver, to construct solutions to constraint problems. Z3 is very expressive and can handle circular constraints and nonlinear inequalities. This expressiveness leads to some of the low-viscosity properties of Basalt's design. However, nearly all of the constraints used to create the \textit{DFSCQ} diagram are sparse linear equations similar to the ones shown in \Cref{fig:basaltvbluefish}. As a result, it may be possible to achieve the same functionality with a local propagation system like Bluefish's.
    



\section{Design Reflections with a Professional Creative Coder}
\label{sec:reflection}
We built the example gallery in collaboration with a professional creative coder, Elliot Evans. Except for the Python Tutor diagram, Evans directed the implementation of the examples. Through discussions during and after building these examples, we surfaced two main insights about Bluefish's abstractions.

\subsection{Relations provide a shallow learning curve for UI developers}
In addition to providing conceptual simplicity, relaxing the component model makes Bluefish easier to understand for UI developers. Evans found that programming in Bluefish without using relations was very similar to using Tailwind. For example, consider Tailwind's \texttt{flex-row} specification versus Bluefish's \texttt{StackH}:

\begin{codeblock}
    \begin{code}
\definecolor{MineShaft}{RGB}{59,59,59}
\definecolor{Maroon}{RGB}{128,0,0}
\definecolor{Red}{RGB}{229,0,0}
\definecolor{Black}{RGB}{0,0,0}
\definecolor{Tamarillo}{RGB}{163,21,21}
\textcolor{Maroon}{<}\textcolor{Maroon}{div}\textcolor{MineShaft}{ }\textcolor{Red}{class}\textcolor{Black}{=}\textcolor{Tamarillo}{"flex flex-row gap-5"}\textcolor{Maroon}{>}\\
\textcolor{MineShaft}{  }\textcolor{Maroon}{<}\textcolor{Maroon}{div}\textcolor{Maroon}{>}\textcolor{MineShaft}{1}\textcolor{Maroon}{</}\textcolor{Maroon}{div}\textcolor{Maroon}{>}\\
\textcolor{MineShaft}{  }\textcolor{Maroon}{<}\textcolor{Maroon}{div}\textcolor{Maroon}{>}\textcolor{MineShaft}{2}\textcolor{Maroon}{</}\textcolor{Maroon}{div}\textcolor{Maroon}{>}\\
\textcolor{MineShaft}{  }\textcolor{Maroon}{<}\textcolor{Maroon}{div}\textcolor{Maroon}{>}\textcolor{MineShaft}{3}\textcolor{Maroon}{</}\textcolor{Maroon}{div}\textcolor{Maroon}{>}\\
\textcolor{Maroon}{</}\textcolor{Maroon}{div}\textcolor{Maroon}{>}
\end{code}
\end{codeblock}

\begin{codeblock}
    \begin{code}
\definecolor{MineShaft}{RGB}{59,59,59}
\definecolor{Maroon}{RGB}{128,0,0}
\definecolor{JellyBean}{RGB}{38,127,153}
\textcolor{Maroon}{<}\textcolor{JellyBean}{StackH}\textcolor{Maroon}{>}\\
\textcolor{MineShaft}{  }\textcolor{Maroon}{<}\textcolor{JellyBean}{Text}\textcolor{Maroon}{>}\textcolor{MineShaft}{1}\textcolor{Maroon}{</}\textcolor{JellyBean}{Text}\textcolor{Maroon}{>}\\
\textcolor{MineShaft}{  }\textcolor{Maroon}{<}\textcolor{JellyBean}{Text}\textcolor{Maroon}{>}\textcolor{MineShaft}{2}\textcolor{Maroon}{</}\textcolor{JellyBean}{Text}\textcolor{Maroon}{>}\\
\textcolor{MineShaft}{  }\textcolor{Maroon}{<}\textcolor{JellyBean}{Text}\textcolor{Maroon}{>}\textcolor{MineShaft}{3}\textcolor{Maroon}{</}\textcolor{JellyBean}{Text}\textcolor{Maroon}{>}\\
\textcolor{Maroon}{</}\textcolor{JellyBean}{StackH}\textcolor{Maroon}{>}
\end{code}
\end{codeblock}

Evans first familiarized himself with Bluefish by using it as a UI layout engine only, without \texttt{Ref}. He then learned Bluefish's relations concept through a bridge example much like our example in \Cref{fig:ui-vs-bluefish}. Specifically, he started building the \textit{Quantum Circuit Equivalence} diagram solely using nested hierarchies before placing the \texttt{Background} highlight using a \texttt{Ref}. After using \texttt{Background}, a traditional UI component, in an overlapping context, Evans used the \texttt{Line} relation, which has no direct UI component analog.

\subsection{Bluefish specifications often move from hierarchical to diffuse}
While Bluefish specifications often start like UI specifications\dash{}compact and hierarchical\dash{}Evans observed that his diagram specifications typically became more diffuse and relational over time. We demonstrate this pattern in \Cref{sec:tradeoff}.

Through language design conversations with Evans, we realized this behavior stems from our choice to unify different Gestalt relations with a shared abstraction. Relations can be composed to create new elements, which means relations must have bounding boxes. But while some relations (e.g., \texttt{Background}) have bounding boxes that are easy to define, the bounding boxes of other relations like \texttt{Arrow} are more ambiguous. Should \texttt{Arrow}'s bounding box contain the bounding boxes of the elements it connects? To facilitate easy and predictable switching between different relations, we decided that all relations' bounding boxes should contain their children. While this approach lowers editing viscosity, it requires users to denest specifications earlier than they may expect.

For example, consider the \textit{Python Tutor} diagram. It depicts program state of a running Python program. In Bluefish we construct a top-level element that accepts a description of the stack and heap:

\begin{codeblock}
    \begin{code}
\definecolor{MineShaft}{RGB}{59,59,59}
\definecolor{Maroon}{RGB}{128,0,0}
\definecolor{JellyBean}{RGB}{38,127,153}
\definecolor{Red}{RGB}{229,0,0}
\definecolor{Black}{RGB}{0,0,0}
\definecolor{Tamarillo}{RGB}{163,21,21}
\definecolor{Blue}{RGB}{0,0,255}
\definecolor{NavyBlue}{RGB}{0,16,128}
\definecolor{Dallas}{RGB}{121,94,38}
\definecolor{Salem}{RGB}{9,134,88}
\textcolor{Maroon}{<}\textcolor{JellyBean}{PythonTutor}\\
\textcolor{MineShaft}{  }\textcolor{Red}{stack}\textcolor{Black}{=}\textcolor{Blue}{\{}\textcolor{Black}{[}\\
\textcolor{Black}{    \{ }\textcolor{NavyBlue}{variable}\textcolor{NavyBlue}{:}\textcolor{Black}{ }\textcolor{Tamarillo}{"c"}\textcolor{Black}{, }\textcolor{NavyBlue}{value}\textcolor{NavyBlue}{:}\textcolor{Black}{ }\textcolor{Dallas}{pointer}\textcolor{Black}{(}\textcolor{Salem}{0}\textcolor{Black}{) \},}\\
\textcolor{Black}{    \{ }\textcolor{NavyBlue}{variable}\textcolor{NavyBlue}{:}\textcolor{Black}{ }\textcolor{Tamarillo}{"d"}\textcolor{Black}{, }\textcolor{NavyBlue}{value}\textcolor{NavyBlue}{:}\textcolor{Black}{ }\textcolor{Dallas}{pointer}\textcolor{Black}{(}\textcolor{Salem}{1}\textcolor{Black}{) \},}\\
\textcolor{Black}{    \{ }\textcolor{NavyBlue}{variable}\textcolor{NavyBlue}{:}\textcolor{Black}{ }\textcolor{Tamarillo}{"x"}\textcolor{Black}{, }\textcolor{NavyBlue}{value}\textcolor{NavyBlue}{:}\textcolor{Black}{ }\textcolor{Tamarillo}{"5"}\textcolor{Black}{ \}}\textcolor{Black}{]}\textcolor{Blue}{\}}\\
\textcolor{MineShaft}{  }\textcolor{Red}{heap}\textcolor{Black}{=}\textcolor{Blue}{\{}\textcolor{Black}{[}\\
\textcolor{Black}{    }\textcolor{Dallas}{tuple}\textcolor{Black}{(}\textcolor{Tamarillo}{"1"}\textcolor{Black}{, }\textcolor{Dallas}{pointer}\textcolor{Black}{(}\textcolor{Salem}{1}\textcolor{Black}{), }\textcolor{Dallas}{pointer}\textcolor{Black}{(}\textcolor{Salem}{2}\textcolor{Black}{)),}\\
\textcolor{Black}{    }\textcolor{Dallas}{tuple}\textcolor{Black}{(}\textcolor{Tamarillo}{"1"}\textcolor{Black}{, }\textcolor{Tamarillo}{"4"}\textcolor{Black}{),}\\
\textcolor{Black}{    }\textcolor{Dallas}{tuple}\textcolor{Black}{(}\textcolor{Tamarillo}{"3"}\textcolor{Black}{, }\textcolor{Tamarillo}{"10"}\textcolor{Black}{)}\textcolor{Black}{]}\textcolor{Blue}{\}}\\
\textcolor{MineShaft}{  }\textcolor{Red}{heapArrangement}\textcolor{Black}{=}\textcolor{Blue}{\{}\textcolor{Black}{[}\\
\textcolor{Black}{    [}\textcolor{Salem}{0}\textcolor{Black}{, }\textcolor{Blue}{null}\textcolor{Black}{, }\textcolor{Blue}{null}\textcolor{Black}{],}\\
\textcolor{Black}{    [}\textcolor{Blue}{null}\textcolor{Black}{, }\textcolor{Salem}{1}\textcolor{Black}{, }\textcolor{Salem}{2}\textcolor{Black}{]}\textcolor{Black}{]}\textcolor{Blue}{\}}\\
\textcolor{Maroon}{/>}
\end{code}
\end{codeblock}

Each \texttt{pointer} corresponds to an \texttt{Arrow} in the diagram. We initially wanted to place the \texttt{Arrow} relations corresponding to stack pointers inside the stack element. This nesting would mirror the data structure driving the visualization. However, the \texttt{Arrow} relation's bounding box contains the heap object it points to. The \texttt{Arrow} must therefore be denested out of the stack element or else the stack would contain the \texttt{Arrow} and thus the heap object. The tradeoff of this early denesting is that switching the \texttt{Arrow} relation for a \texttt{Background} is a predictable, atomic edit. If the \texttt{Arrow} were nested and its bounding box did not contain its children, the user might be surprised that switching it to a \texttt{Background} would suddenly include a heap object in the stack.

\section{Discussion and Future Work}
\label{sec:discussion}

In this paper, we presented Bluefish, a diagramming framework based on relations that is declarative, composable, and extensible. We have demonstrated how relaxing the component model transfers the benefits of UI framework design to diagramming without the need for completely new concepts like constraints. Relations allow users to smoothly trade the local affordances of hierarchical specification for the expressive affordances of adjacency. Our long-term goal is to make Bluefish both a usable tool and a research platform for investigating graphic representations from diagrams to documents to notation augmentations the way Vega-Lite has done for statistical graphics~\cite{vega-lite} and LLVM for compilers~\cite{lattner2004llvm}. To support this goal, we have released Bluefish as an open source project at \href{https://bluefishjs.org}{bluefishjs.org}, and present several promising directions for future research and tool development.


\textbf{Interactive and Animated Graphic Representations.}
In this paper, we explored formalisms of relations for \emph{static} graphics. An immediate next step would be to consider how our abstractions could be extended to interactive and animated diagrams.
First, there are temporal analogs to static Gestalt relations. For example, \textit{common fate}, where elements travel in the same direction are grouped together, is alignment applied to velocity~\cite{gestalt}.
%
%
Similarly, we could think of Bluefish's \verb|Distribute| as distributing elements along a time axis to stagger movements in time, and a temporal \verb|Align| as unifying the start or end of multiple animations. 
Data visualization grammars have explored these temporal analogs. For example Gemini provides \verb|concat| and \verb|sync| operations for temporal distribution and alignment, respectively~\cite{kim2020gemini}. 
Animations may also be staged or nested, conveying information similar to \emph{common region}, as in Canis/CAST~\cite{ge2020canis,ge2021cast}. 
Or an animation may follow a path between two elements to represent a temporal \emph{link} between them.
There are analogs in interaction as well. Gestalt relations seem to manifest in interactions as \textit{on-demand} relations. For example, brushing can be thought of as on-demand common region,
and generalized selections~\cite{heer2008generalized}
allow users to select sets of elements based on \textit{similar attributes}.

\textbf{Formalizing Visual Structure and Domain Semantics.}
%
%
While our standard library of relations covers a large number of use cases, many domains have different sets of primitive relations.
%
For example,
Euclidean geometry features relations like line-line intersection and perpendicular bisector. Similarly, specific features of a line convey semantic intent in sketched route maps~\cite{tversky2000lines}. A straight line means \textit{``go down,''} a curved line means \textit{``follow around,''} and a line with a sharp corner can signify a \textit{``turn.''} These primitives tie closely to the underlying semantics of the domains they visualize, synthetic planar geometry and routes, respectively. Mackinlay's expressiveness principle\cite{mackinlay1986automating}, Tversky's correspondence principle~\cite{tversky2019mind}, and Kindlmann and Scheidegger's algebraic design process~\cite{kindlmann2014algebraic} suggest we may find many such mappings between graphics and domain semantics.
The core idea underlying Bluefish is that more powerful formalisms of these correspondences not only lower authoring viscosity, but also capture more underlying semantic information for later analysis and processing.

\textbf{Towards Richer Tools for Graphic Representations.}
Developing these formal mappings also enables more powerful tools for end-users.
For instance, how might Bluefish's scenegraph\dash{}which explicitly encodes relationships between elements\dash{}be automatically retargeted for screen reader use, blending approaches found in tools such as Olli~\cite{blanco2022olli}, which produces a hierarchical structure for navigating statistical graphics, and Data Navigator~\cite{elavsky2023data}, which provides methods for navigating adjacency structures?
%
Similarly, while diagramming environments such as StickyLines~\cite{ciolfi2016beyond} have reified alignment and distribution, Bluefish's relations suggest the possibility of a more general, consistent interface for allowing end-users to directly manipulate Gestalt relations.
Tools like Draco~\cite{2019-draco} and Scout~\cite{scout} have explored automatic recommendations of statistical graphics and UIs, respectively, based on studies from the perceptual literature. By providing an explicit encoding of relations at the language level, we believe Bluefish can serve as the base for exploring diagramming recommendations based on the relative effectiveness of Gestalt relations.

\begin{acks}
Thanks to Josh Horowitz for pointing us to modern UI local propagation layouts; to Tom George for his help on a previous iteration of the project; and to our anonymous reviewers for their thoughtful feedback that clarified our contributions. This work is supported by the National Science Foundation under Grant No. 1745302.
\end{acks}

\bibliographystyle{ACM-Reference-Format}
\bibliography{main}

\end{document}